\documentclass[sigconf,edbt]{acmart-edbt2020}
\newif\iffinal
\finaltrue
\usepackage{amsfonts}
\usepackage{textcomp}
\usepackage{hyperref}
\usepackage{graphicx}
\usepackage{alltt}
\usepackage{xcolor}
\usepackage{balance}
\usepackage{microtype}
\usepackage{amssymb}
\usepackage{amsmath}
\usepackage{algorithm,algorithmic}
\usepackage{mathtools}
\usepackage{enumitem}
\usepackage{subcaption}

\newfloat{algorithm}{t}{lop}
\hypersetup{colorlinks, citecolor=black, filecolor=black, linkcolor=black,urlcolor=blue}

\usepackage{booktabs} % For formal tables

\newcommand{\ns}[1]{{{#1}}}

\iffinal
\newcommand{\reminder}[1]{}
\else
\newcommand{\reminder}[1]{{\color{red}{[[**#1**]]}}}
\fi

\newcommand{\eat}[1]{}

\iffinal
\newcommand{\fullversion}[1]{#1}
\newcommand{\short}[1]{}
\newcommand{\shorttab}[1]{}
\else
\newcommand{\fullversion}[1]{}
\newcommand{\short}[1]{\ns{#1}}
\newcommand{\shorttab}[1]{#1}
\fi

\newcommand{\smalltt}[1]{{\small{\texttt{#1}}}}

\newcommand{\tab}{\hspace*{2mm}}

\usepackage{lipsum}

\newcommand\blfootnote[1]{%
  \begingroup
  \renewcommand\thefootnote{}\footnote{#1}%
  \addtocounter{footnote}{-1}%
  \endgroup
}

\fullversion{
\settopmatter{printacmref=false} 
\renewcommand\footnotetextcopyrightpermission[1]{} 
\pagestyle{plain}
}

\begin{document}
\short{
\title{Edit Based Grading of SQL Queries\thanks{Work done while all authors were at IIT 
Bombay}}}

\fullversion{
\title{Edit Based Grading of SQL Queries}
}

\fullversion{
\author{Bikash Chandra*$^\dagger$}
\affiliation{\institution {EPFL}}
\email{bikash.chandra@epfl.ch}

\author{Ananyo Banerjee*}
\affiliation{\institution {Oracle India}}
\email{94ananyo@gmail.com}

\author{Udbhas Hazra*}
\affiliation{\institution{Apple India}}
\email{udbhas.hazra@gmail.com}

\author{Mathew Joseph*}
\affiliation{\institution {Raymour \& Flanigan Furniture and Mattresses}}
\email{mathew\_joseph31@yahoo.com}

\author{S. Sudarshan}
\affiliation{IIT Bombay}
\email{sudarsha@cse.iitb.ac.in}
}

\short{
\author{Bikash Chandra\inst{1}$^\dagger$ \and
Ananyo Banerjee\inst{2} \and Udbhas Hazra\inst{3} \and Mathew Joseph\inst{4} \and
S. Sudarshan\inst{5}}

\authorrunning{B. Chandra et al.}

\institute{EPFL \email{bikash.chandra@epfl.ch} \and 
Oracle India \email{94ananyo@gmail.com}\and 
Apple India \email{udbhas.hazra@gmail.com}
\and Raymour \& Flanigan Furniture and Mattresses \email{mathew\_joseph31@yahoo.com}
\and IIT Bombay
\email{sudarsha@iitb.ac.in}}

\maketitle
}

\fullversion{
\renewcommand{\shortauthors}{}
}

\begin{abstract}
Grading student SQL queries manually is a tedious and error-prone process.
Earlier work on testing correctness of student SQL queries, such as the XData system, can 
be used to test correctness of a student query.
However, in case a student query is found to be incorrect there is currently no way to 
automatically assign partial marks. Partial marking is important so that small errors are 
penalized less than large errors. Manually awarding partial marks is not 
scalable for classes with large number of students, especially MOOCs, and is also prone 
to human errors. 

In this paper, we discuss techniques to find a minimum cost set of edits to a student 
query that would make it correct, which can help assign partial marks, and to help 
students understand exactly where they went wrong. Given the limitations of 
current formal methods for checking equivalence, our approach is based on finding nearest 
query, from a set of instructor provided correct queries, that is found to be equivalent 
based on query canonicalization. We show that exhaustive techniques are expensive, and 
propose a greedy heuristic approach that works well both in terms of runtime and accuracy 
on queries in real-world datasets. Our system can also be used in a learning mode where 
query edits can be suggested as feedback to students to guide them towards a correct 
query. Our partial marking system has been successfully used in courses at IIT 
Bombay and IIT Dharwad.
\blfootnote{
\fullversion{
\hspace*{-0.07cm}$^*$Work done while all authors were at IIT Bombay\\
$^\dagger$ Work partially supported by a PhD fellowship from Tata Consultancy Services}
\short{
\hspace*{-0.57cm}$^\dagger$ Work partially supported by a PhD fellowship from Tata 
Consultancy Services
}}
\end{abstract}

\fullversion{
\maketitle
}

\section{Introduction}
\label{sec:introduction}

Grading SQL queries is typically done either by manually checking whether the SQL 
query submitted by the student matches the correct query or by comparing results of 
student SQL queries with that of correct SQL queries on one or more fixed datasets. 
Manual checking of SQL queries is cumbersome and error-prone. \fullversion{Consider a 
case where the correct query provided by the instructor is 
\\\smalltt{\tab SELECT id, course\_id \\\tab FROM student LEFT OUTER JOIN (SELECT * FROM 
takes
 \\\tab\tab  WHERE takes.year=2018) USING(id)\\ 
}
while the student submits the query \\
\noindent\smalltt{\tab SELECT id, course\_id \\\tab FROM  student LEFT OUTER JOIN takes
USING(id) \\\tab WHERE takes.year=2018
}

The query submitted by the student is not equivalent to the one provided by the 
instructor since it does not output students who have not taken a course in 2018. 
The instructor's query outputs these tuples with a null value for course\_id. A grader
evaluating the student query may miss the difference.

Fixed datasets are query agnostic and when used for evaluating student 
queries may fail to catch errors in student SQL queries. For the above example, a 
dataset can catch the error only if it has a tuple $s1$ in the student relation such that 
(i) there is a tuple in the takes relation with $id=s1.id$ and (ii) no tuple in the takes 
relation corresponding to that $id$ has $year=2018$. 

}
The XData~\cite{xdata:vldbj15,xdata:icde11} system generates one or more datasets to 
catch common errors in SQL queries. The datasets generated are tailored to catch errors 
of a particular query. Datasets generated by the XData system based on instructor queries 
can hence be used to execute the instructor and student queries and compare the 
results. We give a brief background of data generation in Section~\ref{sec:background}. 
\ns{There are other techniques for checking query equivalence but, as discussed in 
Section~\ref{sec:relwork}, they have limitations in terms of SQL features handled.}
However, for grading, just 
detecting that a query is incorrect may not be sufficient; it is also 
necessary to provide partial marks to incorrect SQL queries in such a way that small 
errors are penalized less than major errors. 

A naive approach could be to 
award partial marks based on the fraction of datasets where 
the results of instructor query and student query match. However, this approach gives 
very poor results since almost correct student queries may get penalized heavily 
for a small error while student queries that do not even use the correct tables may get 
some marks. \fullversion{We show such examples in Section~\ref{sec:bg:naive}. \eat{The 
approach may also not help student understand their mistakes.}}

\ns{

When evaluating student queries, a grader typically manually identifies changes 
that are required in the student query to make it equivalent to a correct query. 
Consider a correct query provided by the instructor to be 
\\ \smalltt{
\tab SELECT * FROM r INNER JOIN s ON (r.A=s.A)
WHERE r.A$>$10
}
\\ Consider a student query 
\\ \smalltt{
\tab SELECT * FROM r INNER JOIN s ON (r.A=s.B)
WHERE s.A$>$10\\
} 
A grader 
evaluating the student query above may deduct marks for two errors - one for the join 
condition 
and another for the selection condition. However, if the join condition in the student 
query is fixed, the student query is equivalent to the given correct query since now 
\smalltt{r.A} and \texttt{s.A} are equivalent in the student query. Hence only marks for 
one error should have been deducted.
}

\ns{In this paper, we discuss techniques to find the minimum cost sequence of edits to an 
incorrect student query that would make it correct. This is used to automatically award 
partial marks based on the required changes and is also useful in helping students 
understand exactly the mistakes in their query.}
Awarding marks by identifying such changes required in the student query allows us to 
award partial marks in a calibrated manner; a student query that needs 
more changes can be awarded less marks compared to a student query that needs less 
changes. However, checking if the edited student query is equivalent to a correct query 
is difficult. \ns{As mentioned earlier, checking for semantic equivalence is non-trivial 
for SQL queries in general. Checking for equivalence using syntactic identicalness is too 
strong a condition.} Many syntactic differences cause no difference in the query result. 
For example, the selection condition 
\smalltt{r.A$>$5} may also be written as \smalltt{5$<$r.A}; we use query canonicalization 
to replace selection conditions with 
\smalltt{$>$} to equivalent conditions with \smalltt{$<$}. 
Similarly, \smalltt{ORDER BY student.id, student.name} can also be written as 
\smalltt{ORDER BY student.id}, since \smalltt{student.id} determines 
\smalltt{student.name}.
We use a variety of query canonicalization techniques to remove many irrelevant syntactic 
and semantic differences between 
the student and instructor queries and then compute the edit distance between them. If 
the edit distance after canonicalization (which we call \textit{canonicalized edit 
distance}) is 0, the queries are identical. While canonicalized edit distance between 
student and correct queries could directly be used for partial marking, we show that it 
has some limitations.

In this paper, we instead propose techniques to award partial marks to a student SQL 
query based on the query edits required to make it equivalent to a correct query provided 
by the instructor. 
The edits could be in the form of insertion, deletion, replacement or movement of parts 
of the query.  The weight of each type of edit can be changed by the instructor if 
desired. 
The instructor may provide multiple correct queries. We compute partial marks for the 
student query with respect to all correct queries and choose the best match  i.e. the 
one that gives the highest marks.
 
Our contributions in this paper are as follows. 
\begin{enumerate}
\item \ns{We have developed canonicalization techniques which 
enable us to reduce irrelevant differences between the student and correct 
queries. Once the queries are canonicalized, we can find the edit distance between the 
student query and correct query which we call canonicalized edit distance.
Since prior work on techniques for checking for equivalence do not handle many 
SQL features, we use canonicalized edit distance equals 0 as a sufficient condition for 
equivalence testing.

The canonicalization techniques not only take care of irrelevant syntactic 
differences between the student query and a correct query but also take into account 
database constraints (such as foreign keys) and query constraints (such as query 
 predicates). The canonicalization techniques are discussed in 
 Section~\ref{sec:canonicalize}. 
Our canonicalization techniques are extensible; new canonicalization rules can be easily 
added to our framework. }

\item The canonicalized edit distance between the student query and the 
correct query may be used to award partial marks. 
However, canonicalized edit distance has drawbacks as discussed in 
Section~\ref{sec:canonical:marks}. 
Hence, we use a query editing based model which makes 
successive edits to student queries to make it equivalent to a correct query. 
\ns{We can then use the edits to identify the mistakes students made or to award partial 
marks.} 
The distance computed based on the edits is called the \textit{weighted edit sequence
distance}. 

\item We show that the problem of finding the weighted edit sequence distance
can be reduced to the problem of finding the 
shortest path in a graph. Since the graph is dynamically generated and potentially very 
large, we also describe a greedy heuristic technique for finding the 
weighted edit sequence distance that uses the 
canonicalized edit distance to prune the search space. The heuristic performs 
well both in terms of runtime and accuracy in real datasets. We discuss these techniques 
in Section~\ref{sec:edit_seq}.
Our query edit rules are also extensible. 

\item In Section~\ref{sec:expt}, we describe our experimental results performed over 
student queries \fullversion{collected} from an undergraduate course at IIT Bombay from 
2015 to 2017. 
We show the effectiveness of our techniques in terms of fairness of marks awarded as well 
as execution time. \emph{Our partial marking techniques have also been used in evaluating 
queries in database courses at IIT Bombay and IIT Dharwad in 2018. \ns{The students 
contested much fewer queries compared to previous years when partial marks were awarded 
manually.}}

\end{enumerate}

In Section~\ref{sec:background}, we briefly discuss some background and define the class 
of queries supported. Related work is discussed in Section~\ref{sec:relwork}. \ns{The 
queries used in this paper are based on the University Schema \cite{dbconcepts2010}.}
\short{ Parts of the canonicalization techniques were published 
earlier as a demo 
\cite{xdata:vldb16}. In this paper, we show why using canonicalization alone does not 
work well for grading student queries. We also discuss additional canonicalization 
techniques including flattening equality predicates. 
A short version of this paper 
has been published as a poster \cite{xdata:icde19}. In this paper, we additionally 
present the space of guided edits we consider along with 
algorithms for finding edit sequence. 
We also present additional experimental results including experiments on effectiveness of 
canonicalization and comparison of greedy and heuristic techniques for finding edit 
sequence. A full version of this paper is available at \cite{arxiv}.} 

\section{Preliminaries}
\label{sec:background}

\begin{figure}
		\centerline{\includegraphics[width=0.5\textwidth,keepaspectratio=true]{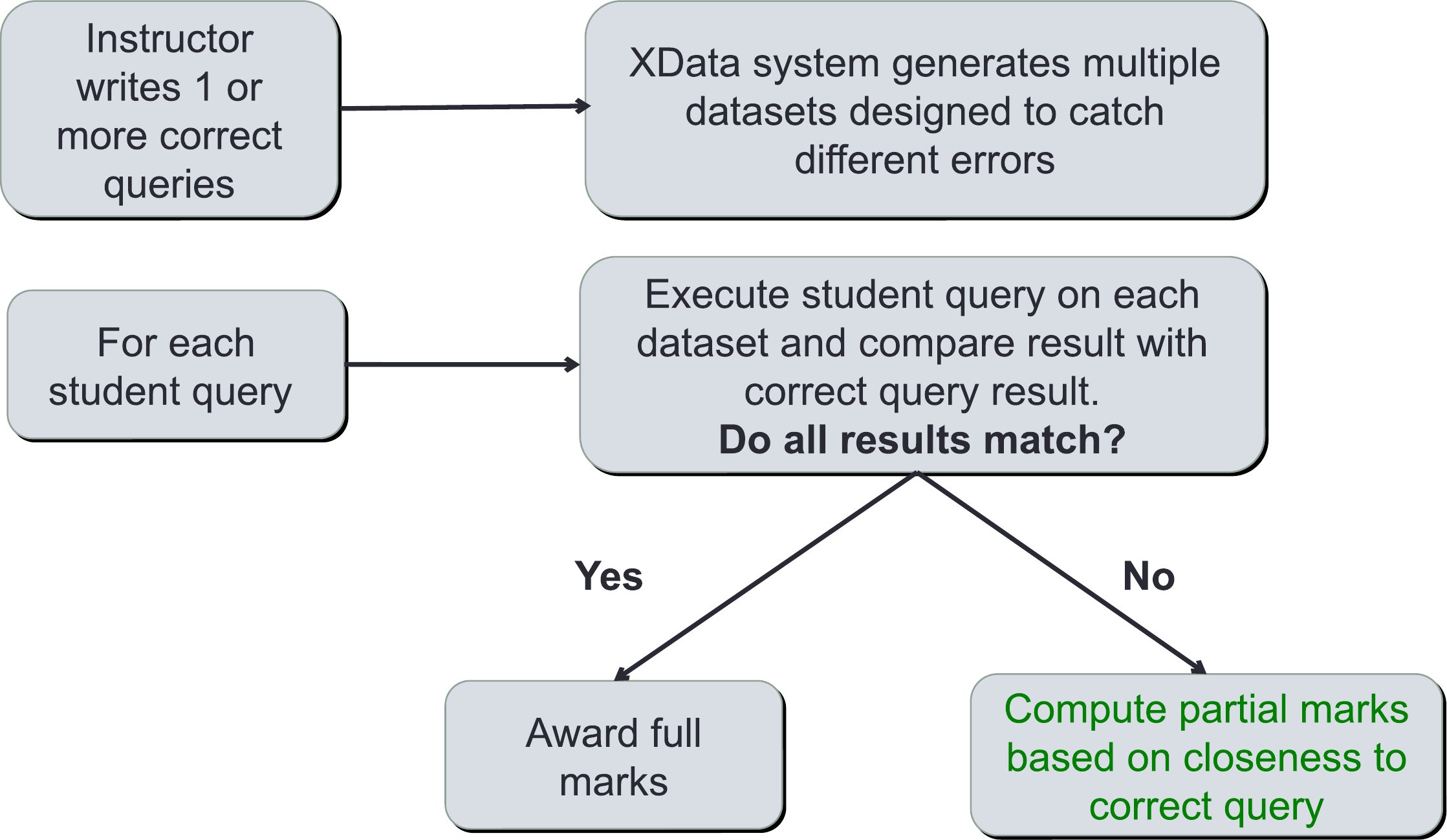}}
		\caption{Automated Grading Workflow}
		\label{fig:workflow}
\end{figure}

In this section, we give a brief overview of the XData system for data 
generation\fullversion{ and 
show why datasets cannot be used for partial marking}. We also give a list of the class 
of queries that we support for partial marking. 
The overall workflow of our grading system is shown in Figure~\ref{fig:workflow}.

\subsection{Checking Query Correctness Using XData}
\fullversion{Incorrect student queries are often deviations (or mutations) of correct 
queries. A 
mutation is defined as a single syntactically correct change to the correct 
query and the changed query is said to be a mutant of the original query. A dataset that 
is able to produce different results on the correct query and its mutant (thereby showing 
that the mutant is not equivalent to a correct query) is said to kill the mutation.

The XData~\cite{xdata:vldbj15,xdata:icde11} system takes a query as input and 
generates one or more datasets such that common errors/mutations are killed by one of 
the datasets. 
The XData system handles a large variety of SQL constructs including selections, joins, 
aggregates, subqueries and set operators. Currently, the mutations targeted by XData 
includes join mutations, comparison operator mutations, aggregate operator mutations, 
group by attribute mutations, like mutations, subquery mutations, set operator mutations 
and distinct mutations. For each input query, XData generates multiple datasets, each 
targeted to kill one or more mutations. 
}

\short{The XData~\cite{xdata:vldbj15,xdata:icde11} system takes a query as input and 
generates multiple datasets to catch common errors in the given query.
The XData system handles a large variety of SQL constructs including selections, joins, 
aggregates, subqueries and set operators.}

For the purpose of grading, we use one or more correct queries provided by instructors to 
generate datasets. These datasets can be used to compare the result of student query and 
instructor query to allow killing of non-equivalent mutations in the student query. If 
the result of the student query matches that of the instructor query on all the generated 
datasets, the student query is marked as correct and full marks are awarded to it. 
(Correctness is not guaranteed, but this is a best effort test.) If 
there is a mismatch in the student query result and the instructor query result on any 
dataset, the query is marked as incorrect and no marks are awarded. Instructors or TAs 
need to manually go over the incorrect queries and award partial marks, which is 
time-consuming, error-prone and does not scale.

\fullversion{
\subsection{Dataset Based Partial Marking}
\label{sec:bg:naive}
A naive way of partial marking, based on generated test data, would be to award marks 
based on the fraction of datasets passed by the student query. However, this may not be 
fair to the students. Consider, the case where the correct query provided by the 
instructor is
~\\ \smalltt{
\tab SELECT id,name \\\tab FROM student INNER JOIN takes 
 USING(id) \\\tab WHERE year$>$2016
}

Suppose a student query incorrectly used \smalltt{year$<$2016} in place of 
\smalltt{year$>$2016} (as specified by the correct query). This student query would fail 
almost all datasets, giving very poor marks. On the other hand, another student query 
 \\ \smalltt{
\tab SELECT dept\_name,building \\\tab  FROM department WHERE FALSE
}\\
would pass datasets that produce 
an empty result on the instructor query. A student submitting this query would get some 
marks without even using the correct set of tables. Hence, it may not be desirable to 
award partial marks using the fraction of datasets passed. 
}

\subsection{Class of Queries}
We consider the following class of queries for partial marking. 
\begin{enumerate}
 \item Single block queries with join/outer-join operations 
and predicates in the where-clause, and optionally aggregate operations, 
corresponding to select / project / join / outer-join queries 
in relational algebra, with aggregation operations.
\item Multi-block queries with nested subqueries in select / from / where clause, which 
may 
have arbitrary levels of nesting.\footnote{Data generation techniques in XData can 
currently handle only one level of nesting in \smalltt{WHERE} clause subqueries.}
\item Compound queries with set operators UNION(ALL), INTERSECT(ALL) and EXCEPT(ALL). 
\end{enumerate}

Our canonicalization and edit techniques work on the parsed query trees.
We assume that the student and correct queries are well formed enough to construct parse 
trees out of them. 
Our system gives immediate feedback when syntactically incorrect queries are submitted.
\ns{When finding edits, we do not penalize variation in column names of the student 
query. 
 }

\section{SQL Query Canonicalization}
\label{sec:canonicalize}

The same SQL query can be written in multiple correct ways. Our canonicalization 
techniques aim to transform queries so that they can be made comparable as a sufficient 
check for equivalence and to ignore irrelevant syntactic differences when computing edit 
distances. 

\fullversion{
\subsection{Motivation}
The key idea of this paper is to successively edit a given student query to make it 
equivalent to a correct query provided by the instructor. \fullversion{This technique 
gives us a 
measure of how much change would have to be made to the student query to make it a 
correct query, making it a fairer way to award partial marks compared to using the 
fraction of datasets passed. There could be many different ways of writing the same query 
and the instructor 
can provide multiple correct queries. We award partial marks to the student query 
against all correct queries provided by the instructor and choose the maximum marks among 
the marks awarded for each query. 

}
We use a set, EQ of edited student queries generated from the given student query, 
initially containing only the given query; 
at each step, we consider further edits to each query in EQ, and add them to EQ to create 
a larger set of edited queries. 
We can stop creating edits once we find an edited query that is equivalent to a correct 
query.
There are two major challenges.
\begin{enumerate}
\item We need to test for equivalence between an edited query and correct query after 
each edit step. 
\item If all possible directed edits at each step are considered for further processing, 
the search space would be exponential in terms of the number of edits required. 
\end{enumerate}

One way to test for equivalence could be to use datasets generated by XData. However, it 
would be very expensive to load each dataset and check for equivalence after each edit.  
Other techniques for checking equivalence such as Cosette~\cite{cosette} and techniques 
based on tableau~\cite{tableaux}, \cite{tableaux1}, \cite{tableaux2}
 work on a limited set of queries. These techniques also 
do not provide an efficient way of pruning the search space.

Canonicalization provides a sufficient condition to establish 
equivalence of queries for the first challenge above. 
In practice, we found that using canonicalization as a 
sufficient test for equivalence works well.
The edit distance computed after canonicalization can 
also be used as a guidance heuristic for prioritizing edits (Section~\ref{sec:heuristic}).

}

\subsection{Query Canonicalization Rules}

In order to reduce irrelevant syntactic differences, we do some initial preprocessing. 
These steps replace certain SQL operators with other operators, 
enabling us to reduce the number of types of operators we need to consider during 
comparison of student and correct queries. For example, \smalltt{NOT(A>B)} can 
be replaced by \smalltt{A<=B}. Some operators like inner joins (provided the join inputs 
do not have DISTINCT, GROUP BY or aggregations) are associative and commutative. We 
flatten such operators and construct a flattened tree  as shown in 
Section~\ref{sec:flattened_tree}. 
\fullversion{We provide a list of syntactic canonicalizations in 
Appendix~\ref{sec:canonical_basic}.} 

Differences between student and correct queries can also be reduced by semantic 
canonicalizations. 
These canonicalizations can only be applied to queries provided some query and/or 
database constraints are satisfied. 
Some of these transformations are widely used in 
query optimizers to consider alternative query execution plans.
However, as shown in \cite{xdata:vldbj15}, using the query plans provided by the 
PostgreSQL database engine to directly compare student and correct queries performs very 
poorly. We use the transformations to make a correct query comparable to a student query.
\short{These canonicalizations include distinct removal, join type transformation and 
order by as well as group by attribute transformation. }

\fullversion{
Some of the semantic canonicalization rules can be summarized as follows.
\begin{itemize}
\item Distinct clauses are removed where possible based on primary keys and unique 
attributes.
\item Outer joins are converted to inner joins in case there is a null rejecting 
condition above the join or in case the join is based on non nullable foreign key 
attributes.
\item Query predicates are pushed down in the query tree where possible.
\item Conjuncts of predicates of the form $A=B$, $B=C$, $C=D$ form an equivalence 
class and any occurrence of an attribute in an equivalence class is replaced by 
the lexicographically attribute smallest at any place in the query tree that is above 
the predicates. In case there is a constant in the equivalence 
class it is considered to be the lexicographically smallest attribute.

\item  Functional dependencies are used to remove additional attributes in the order by 
clause and to compare attributes of the group by clause.
\end{itemize}

Details are provided in Appendix~\ref{sec:semantic}.}

It should be noted that once a canonicalization has been applied on a query, 
canonicalization rules that were not applicable on the original query may become 
applicable now. Hence, we repeat the process of canonicalization (using both syntactic 
and semantic canonicalization rules) until no further canonicalization rules are 
applicable. \ns{Each canonicalization rule is simple, and it is easy to prove each rule 
correct. A sequence of transformations would thus preserve correctness.}
\short{Our current canonicalization rules are carefully designed to ensure termination 
and confluence. Details of the canonicalization techniques are provided in \cite{arxiv}.}

\fullversion{
Our current canonicalization rules are carefully designed to ensure termination and 
confluence. Our canonicalization rules ensure that once a canonicalization rule has been 
applied to transform a  pattern P1 in the query tree to P2, P1 will not be added again, 
thus ensuring termination. Our canonicalization 
rules also ensure if a canonicalization rule R1 was applicable on a query Q and another 
canonicalization rule R2 is applied to transform Q to Q', then R1 will continue to be 
applicable on Q', thus ensuring confluence.
If new canonicalization rules that do not ensure termination or confluence are to be 
used, techniques from query optimization based on transformation rules like 
Volcano~\cite{volcano} could be used to create a DAG representation of alternatives 
instead of choosing between rules. Unification 
techniques from multi-query optimization, such as the one described in \cite{pyro}, can 
be used to test query equivalence from the DAG representation; we currently do not use 
this option.
}

\subsection{Flattened Tree Structure}
\label{sec:flattened_tree}
In order to compare the student query and a correct query, we use a ``flattened'' tree 
structure to represent the SQL queries. 

\fullversion{
\begin{figure}
	\centerline{\includegraphics[width=0.35\textwidth,keepaspectratio=true]{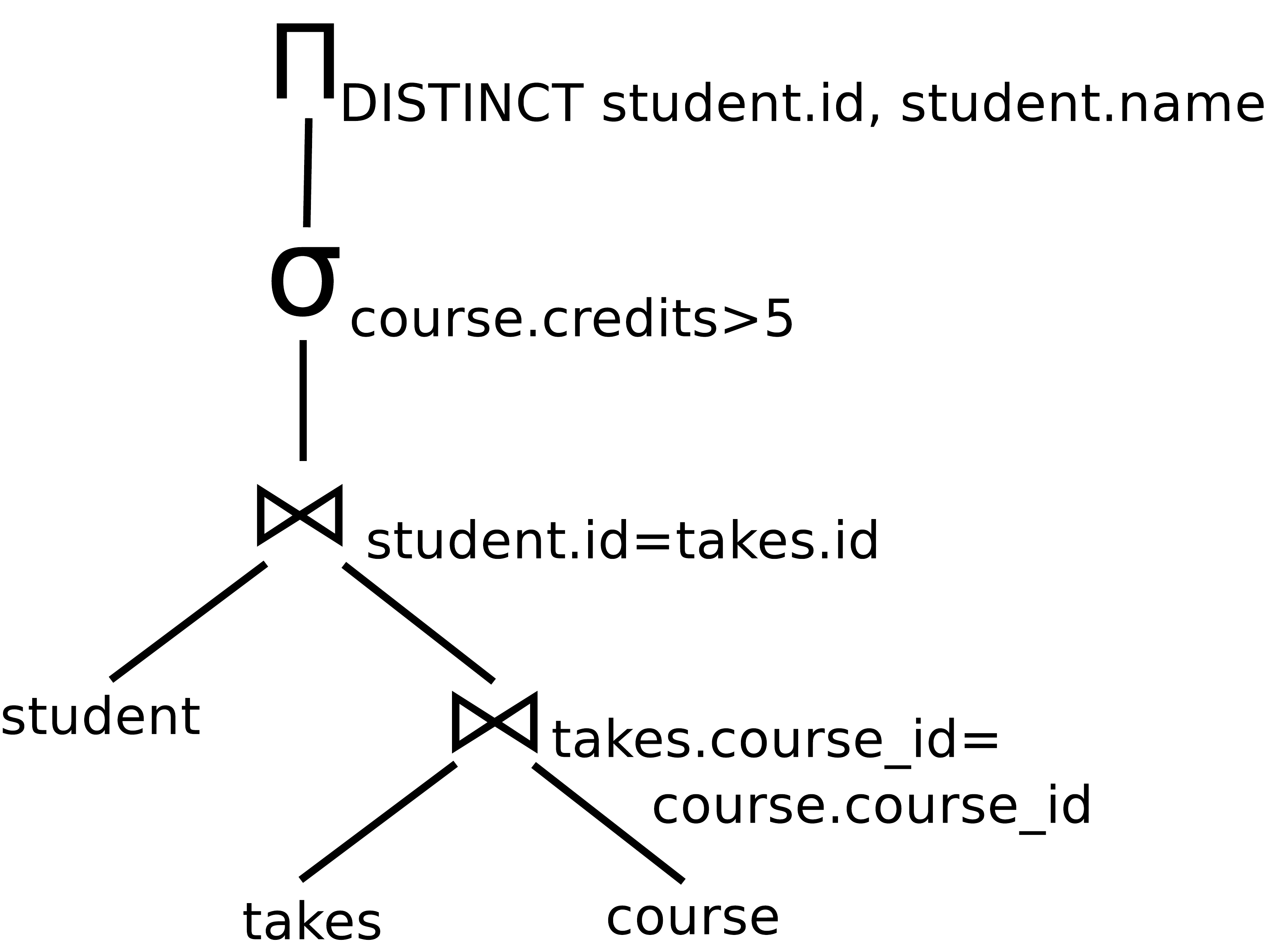}}
	\caption{Parsed Tree}
	\label{fig:parsedTree}
\end{figure}
}

\short{
\begin{minipage}{.45\textwidth}
  \centering
  \includegraphics[width=.9\linewidth]{parsedTree.pdf}
  \captionof{figure}{Parsed Tree}
  \label{fig:parsedTree}
\end{minipage}%
\begin{minipage}{.55\textwidth}
  \centering
  \includegraphics[width=.9\linewidth]{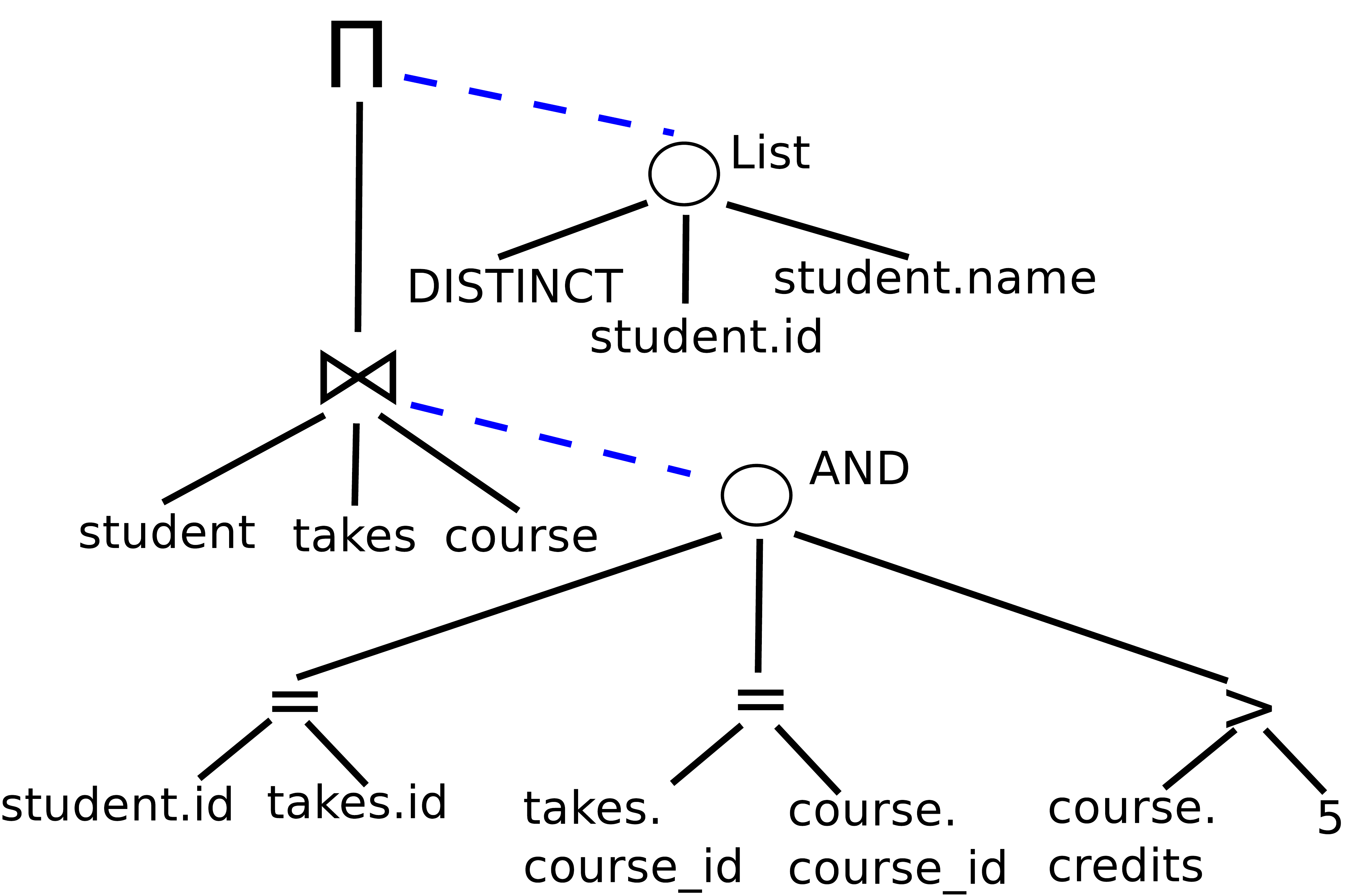}
  \captionof{figure}{Flattened Tree}
  \label{fig:flattenedTree}
\end{minipage}
}

For query operators that are commutative and associative such as INNER JOIN, 
UNION(ALL), INTERSECT(ALL) as well as are predicates involving AND or OR are 
canonicalized by flattening them.\footnote{\ns{We flatten UNION and INTERSECT queries as 
long as the input column names are identical, or the columns are renamed, or not used 
further in the query. }} 
For example $(r\Join s) \Join t $ is transformed to $\Join(r,s,t)$.\footnote{\ns{In case 
the JOIN uses a \texttt{SELECT *}, we explicitly list all attributes in the projection 
list starting from the lexicographically smallest relation.}}
Similarly, for conjuncts of 
predicates with equality which involve common attributes, the attributes form an 
equivalence class and the equality conditions may be specified using different attribute 
combinations. For example,  $(A=B) \wedge (B=C) \wedge (C=D)$ can also be specified as
$(A=B) \wedge (B=C) \wedge (A=D)$ or $(A=B) \wedge (A=C) \wedge (C=D)$.
Regardless of which form is given, the predicate is transformed to $=(A,B,C,D)$.

Consider the query\\
\smalltt{
\tab SELECT DISTINCT student.id, name 
\\\tab FROM student INNER JOIN 
\fullversion{\\\tab \tab }(takes INNER JOIN course USING(course\_id)) USING(id) 
\\\tab WHERE course.credits$>$5
}

\fullversion{
\begin{figure}
	\begin{center}
			\includegraphics[width=0.45\textwidth,keepaspectratio=true]{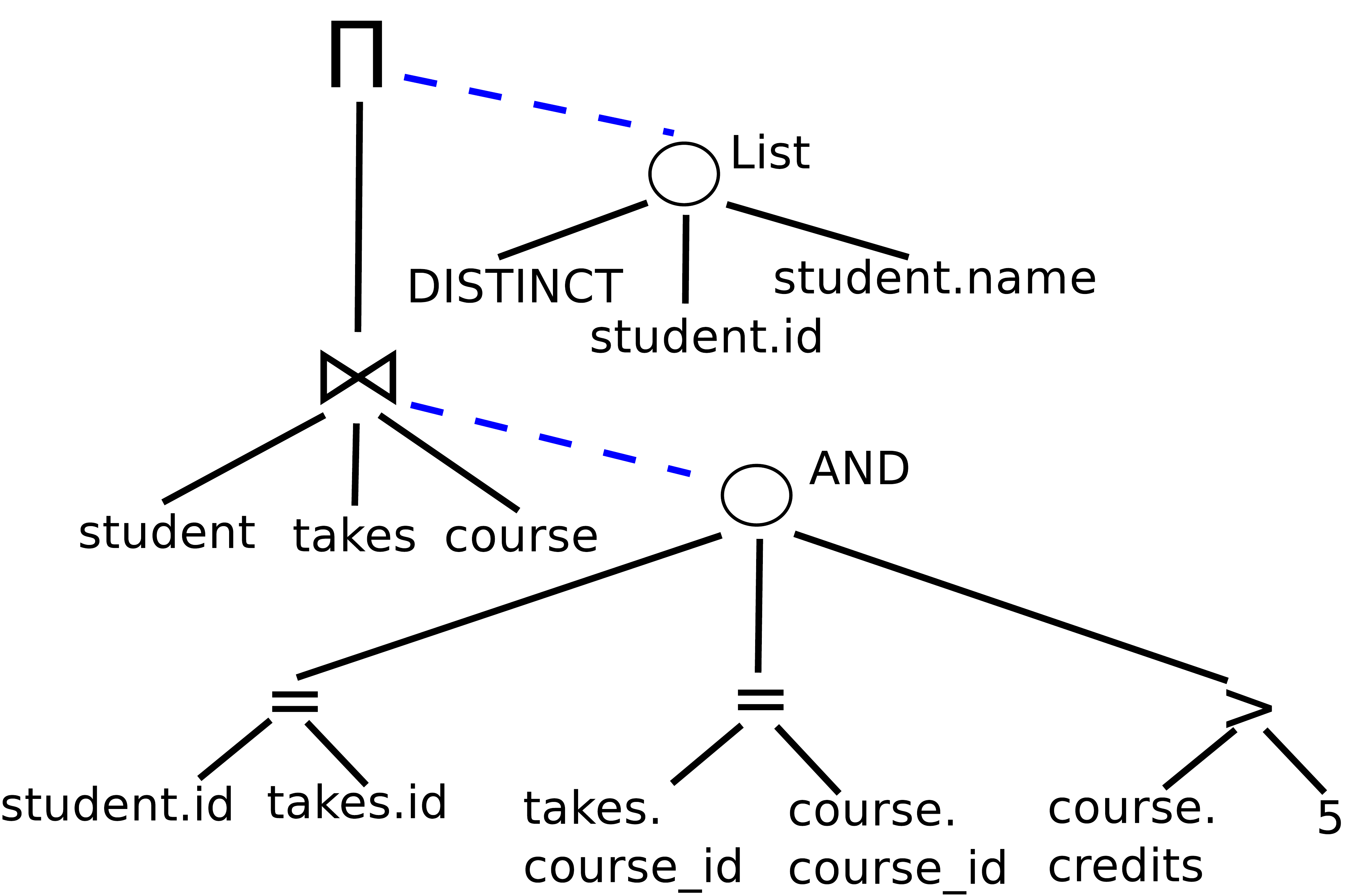}
			\caption{Flattened Tree}
			\label{fig:flattenedTree}
		\end{center}
\end{figure}
}

The parsed tree for the query is shown in Figure~\ref{fig:parsedTree}; the 
flattened tree is shown in Figure~\ref{fig:flattenedTree}.  
Predicates, projections, group by attributes are modeled as special children (connected 
by a dashed line)  may themselves be a subtree.
As shown in Figure~\ref{fig:flattenedTree}, in case the node is an \smalltt{INNER 
JOIN} this child node would contain all the join and selection conditions. 
\short{For non-commutative operators like \smalltt{LEFT OUTER JOIN} and 
\smalltt{EXCEPT(ALL)} we compare 
children in order while for commutative operators like \smalltt{INNER JOIN}, we ignore 
the order 
when matching children.
}

\fullversion{When comparing flattened trees we consider two types of operators nodes. 
\begin{itemize}[leftmargin=*]
\item Operators with ordered inputs: For operators like LEFT OUTER JOIN, EXCEPT (ALL), 
ORDER BY attribute lists, which are non-commutative,  the order of the inputs to the 
operator matters. 
For these operators, we compare the children nodes in order. 
\item Operators with unordered inputs: For other operators like INNER JOIN, flattened 
AND, flattened $=$ 
and UNION(ALL), which are commutative, the order of operators does not matter. For these 
operators, we ignore the order when  
matching children nodes. 
\end{itemize}
}

\subsection{Computing Canonicalized Edit Distance}
\label{sec:edit_distance}

The instructor can set weights for each of the query construct.
The canonical representations of the student query and a correct query can be used to 
compute the weighted edit distance between them. We call this edit distance the  
canonicalized edit
distance and we use this to prune the search space as discussed later in 
Section~\ref{sec:heuristic}.

\fullversion{
\label{sec:edit:type}
Edits on the query tree can be of any of the following types
\begin{itemize}
	\item inserting a node/subtree into the flattened tree
	\item removing a node/subtree from the flattened tree
	\item replacing an existing node/subtree from a flattened tree with another node/subtree in 
	the flattened tree
	\item moving a node/subtree from one position of the flattened tree to another
\end{itemize}
}

Each query edit has a cost associated with it. The cost of editing a subtree within the 
flattened tree is the sum of the cost of all nodes of the subtree. 
In the canonicalized flattened tree, each part of the query such as selection, 
projection, aggregate or subquery is present as a node/subtree. We call each of these 
 parts a component of the query. Each component of the instructor query is 
assigned a weight; weights can be adjusted by the instructor.
We find the edit distance for each component separately and then find weighted edit 
distance. 

\fullversion{
To find edit distance, we compare the flattened trees and find non-matching 
nodes/subtrees to edit the student query to make it 
equivalent to the flattened instructor query tree. 
For nodes whose children are unordered, we do not consider the order of the children when 
computing order and compare all pairs of nodes/subtrees to get the best match.
For nodes that have ordered children, for example ORDER BY clause, the order of the 
nodes/subtrees is also important and edits must bring the parts into the right order.
For such cases, we use standard algorithms that compute the edit distance between 
sequences. 
}

The canonicalized edit distance is computed using the formula
$ \Sigma_{c \in \textrm{components}} W_c*E_c $
where $W_c$ is the weight assigned to a component and $E_c$ is the edit distance for the 
component. \ns{If the canonicalized edit distance is 0, the queries are equivalent.}

{
\subsection{Using Edit Distance for Partial Marks}
\label{sec:canonical:marks}
}

One way to award partial marks could have been to deduct marks based on the canonicalized 
edit distance between the student query and a correct query. 
\eat{However, as discussed in \cite{xdata:icde19,arxiv}, using this technique to award 
partial marks 
may not be fair since canonicalization may increase the edit distance and a small edit 
may greatly reduce the edit distance. This approach also does not provide a minimum cost 
set of edits to make a query correct.}
{However, using this 
technique to award partial marks may not be fair because of the following issues.

\subsubsection*{a) A small edit may greatly reduce the canonicalized edit distance}
\label{sec:canonical:issues:first}
Consider a correct query provided by the instructor to be 
\\ \smalltt{
\tab SELECT * FROM r INNER JOIN s ON (r.A=s.A)
WHERE r.A$>$10
}
\\ Consider a student query 
\\ \smalltt{
\tab SELECT * FROM r INNER JOIN s ON (r.A=s.B)
WHERE s.A$>$10
}

In the above case, finding the canonicalized edit distance would show 
two differences (i) the join condition - the student query uses \smalltt{r.A=s.B} instead 
of \smalltt{r.A=s.A} (ii) the selection condition - the student query uses a selection 
condition on \smalltt{s.A} instead of \smalltt{r.A}. However, if we make 
one change to the query i.e. replace \smalltt{s.B} in the join condition with 
\smalltt{s.A}, the queries would become equivalent now (the selection condition in the 
student query would now be equivalent to the correct query as \smalltt{r.A} becomes 
equivalent to \smalltt{s.A}). 
The student query is just one edit away from a correct query, not 2 as the distance above 
implies.

\subsubsection*{b) Canonicalizations may increase the edit distance}
Consider the case where a correct query provided by the instructor is
\\\smalltt{
\tab SELECT DISTINCT id, name \\
\tab FROM student INNER JOIN takes USING(id)
\fullversion{\\\tab }WHERE takes.semester='Spring'
}

Suppose a student submits the following query which misses the selection condition
\\\smalltt{
\tab SELECT DISTINCT id, name\\ 
\tab FROM student INNER JOIN takes USING(id)
}

In this case, the student has only missed the selection condition and should 
be penalized for one error. However, once canonicalization including redundant join 
elimination and \smalltt{DISTINCT} removal is done, the student query becomes
\\\smalltt{
\tab SELECT id, name FROM student
}

\noindent since the join with \smalltt{takes} is redundant in the student query and 
\smalltt{id} is the primary key of the \smalltt{student}
relation making \smalltt{DISTINCT} redundant.
 Now the difference between 
student and instructor query consists of differences in relations, join operators and join 
conditions and the distinct operator as well. 
The canonicalized edit distance is greater than if the query had not been canonicalized. 

If we first edit the student query to add \smalltt{takes.semester = 'Spring'} and then 
canonicalize the query, the queries would be equivalent. 
}

\section{Minimum Cost Sequence of Edits}
\label{sec:edit_seq}

We now describe our techniques for finding the the lowest cost 
edit sequence.
Our goal is to edit the student query to make it equivalent to a correct query. 
The minimum number of edits or more precisely the least cost edit sequence gives us a 
measure of how far the student query was from a correct query; 
 partial marks can be awarded based on the sum of the cost of the edits, and these edits 
 can help students understand how to make their query correct.
 We then formalize the problem and give an exhaustive 
search algorithm and a greedy heuristic. 

As discussed earlier, an instructor can specify more than one correct query. The 
techniques described in this section are used to evaluate the student query against each 
correct query provided by the instructor. \ns{The lowest cost edit sequence is then used 
to award partial marks and to tell the student how to make their query correct. }

\fullversion{
\subsection{Edit Sequence}
 
To award partial marks to the student query, we make edits to the student query and then 
use the canonicalized edit distance to check if the edited query is equivalent to the 
correct query. 
In general, multiple edits may be needed on the student query to make it equivalent to a 
correct query. 
Different edit sequences may lead to different results.

{Consider the pair of instructor and student query given in 
Section~\ref{sec:canonical:marks}a. }
Let us consider 2 potential edits that can be made on the student query.
\begin{itemize}
\item Changing the selection condition from \smalltt{s.A>10} to \smalltt{r.A>10}. A 
second edit would be needed to change the join condition.
\item Changing the join condition from \smalltt{r.A=s.B} to \smalltt{r.A=s.A}. Now 
\smalltt{r.A} and \smalltt{s.A} belong to the same equivalence class and the selection 
conditions are also equivalent. Hence only one edit in the student query is needed in 
this case.
\end{itemize}

Our goal is to find the edit sequence that has the least cost. 
Algorithms for finding the least cost edit sequence are described in 
Section~\ref{sec:edit_path:algo}.

An alternative to making edits on the student query could have been to edit correct 
queries to make them equivalent to student queries. However, this approach would not be 
fair.
Consider the case where a correct query is \\ 
\smalltt{\tab SELECT DISTINCT id, name \\
\tab FROM student INNER JOIN takes USING(id)
\fullversion{\\\tab }WHERE takes.semester='Spring' \\
}
and the student query is \\
\smalltt{
\tab SELECT id, name FROM student
}

The student has clearly missed several components of a correct query. However, if the 
correct query is edited to remove the selection condition 
\smalltt{takes.semester} \smalltt{='Spring'} the student and correct query would become 
equivalent as shown in Section~\ref{sec:canonical:marks}. Hence the marks 
corresponding 
to only one selection deletion from the correct query would be deducted, which is 
inappropriate. 
}

\subsection{Guided Edits}
When editing a student query, potentially an infinite number of edit options are possible. 
For example, any query predicate can be added as an edit to the student query. 
However, only those edits that make the edited query more similar to the correct query 
 would be useful.
In order to narrow down the search space, 
we edit student queries in a guided manner such that each edit may reduce the 
difference between the student and the correct query. Hence, components of the correct 
query not present in the student query are added to the student query, components of the 
student query that are not present in the correct query are removed and so on. We call 
these edits guided edits. \short{The specific guided edits we consider in our 
implementation  are provided in \cite{arxiv}.}

The specific guided edits we consider in our implementation include %the following
\begin{itemize}
\item Insertion, removal and replacement of projection attributes, group by attributes, 
distinct, aggregates, selection and join conditions. Edits in projection attributes
 of EXISTS and NOT EXISTS subquery need not be considered since they do not 
affect the query result.
\item Joins can be transformed from INNER to OUTER and vice 
versa. When transforming 
from a flattened INNER JOIN to a LEFT OUTER JOIN, we take into account the correct query 
to decide which nodes form the left children and which form the right children of the 
LEFT OUTER JOIN node. Since any RIGHT OUTER JOIN is transformed to LEFT OUTER JOIN during 
syntactic canonicalization we need not take RIGHT OUTER JOIN into account when making 
query edits.
\item Subqueries using the EXISTS connective can be transformed to NOT EXISTS and vice 
versa. Subqueries of IN/NOT IN/ALL/SOME are converted to EXISTS/NOT EXISTS during 
syntactic canonicalization.
\item ORDER BY attributes can be added, deleted or reordered.
\end{itemize}

\fullversion{
\subsection{Edit Generation}
\label{sec:top_bottom}

Potential guided edits on a student query may be found by comparing the flattened trees 
obtained from the student query and a correct query.
If the sub-trees are to be ordered, as in case of ORDER BY attributes, we do an ordered 
comparison for edits, else we do an unordered comparison.
We compare the flattened trees of the student query and the correct query and enumerate a 
list of add, delete, replace or move edits for sub-trees in the 
student query as described below.

When making edits to the flattened query tree, 
we ensure that the query tree is consistent and a syntactically correct query can 
be constructed from the tree. 
For example, consider editing a selection condition in a query \smalltt{r.A>10}. Deleting 
the left-hand side of the selection condition leaves the tree inconsistent. Hence, we do 
not allow such edits. When we add/modify a 
node/subtree, we check whether the attributes in the new node/subtree are visible (based on scope 
rules) at the node/subtree or not. 
Similarly when deleting a node/subtree $st$, we check if the student query has conditions 
above $st$ that depend on $st$. 

Consider the case where the student query has an extra relation \smalltt{r} in the 
flattened join and has a selection condition \smalltt{r.A$>$5} above it. The correct 
query 
does not have \smalltt{r}. Hence deletion of \smalltt{r} from the student query could be 
a potential edit. However, the edit would make the query inconsistent since the selection 
condition \smalltt{r.A$>$5} refers to a relation which does not exist in the query. Hence 
we do not consider the deletion of \smalltt{r} in the current set of edits. However, 
deletion of \smalltt{r.A$>$5} is considered an edit. Once all the dependencies on 
\smalltt{r} such as \smalltt{r.A$>$5} are deleted, deletion of \smalltt{r} as a potential 
edit would be considered. 

Dependencies can be more complex with subqueries and lateral queries. However, the 
dependencies cannot be cyclic since it would be impossible to evaluate a query with 
cyclic dependency.

Since SQL queries are structured by SELECT, FROM, WHERE and  GROUP BY attributes we match the subtrees 
of the student query to of the corresponding node/subtree in the correct query. 
For example, selection conditions in the correct query are compared with selection 
conditions in the student query only.  
Let the list of nodes/subtrees in the student query be $st_s$ and in the correct query be 
$st_c$.
Nodes/subtrees in the $st_s$ that match nodes/subtrees in $st_c$ are not considered for edits. 
We consider the following edits (i) Nodes/subtrees in $st_c$ that are missing in 
$st_s$ can be added to the student query. (ii) Nodes/subtrees 
in $st_s$ that are not present in $st_c$ can be deleted (iii) Nodes/subtrees in $st_s$ 
that do 
not match a node/subtree in $st_c$ can be modified to a node/subtree in $st_c$ that has no matching 
node/subtree in $st_s$ (iv) Nodes/subtrees in $st_s$ that do not match a 
node/subtree 
in $st_c$ but match another node/subtree in the correct query are moved provided the cost 
of move and edit is less than that of insert and delete 
(v) In case an operator is ordered, we also consider edits involving changing the order 
of nodes/subtrees of the operator. 
}

\subsection{Finding Lowest Cost Edit Sequence}
\label{sec:edit_solution}
There 
are several possible edits for the student query. After these edits are applied there may 
be several possible edits on the edited query as well. After applying the edits several 
times the student query can be made equivalent to the correct query. Partial marks can be 
awarded by deducting the sum of the cost of edits made on the student query.

Consider a graph whose nodes are all queries for the given schema. For any query $Q$, 
edits of the query are also nodes in the graph. Let these edited query be connected to  
query $Q$ with an edge whose weight is the edit cost. 
Queries that are canonically equivalent, i.e. their canonical forms are same, are 
connected by 0 cost edges. 
The sequence of edits that has the least cumulative cost can now be determined 
based on the shortest path in this graph from the student 
query node in the graph to a correct query node. Partial marks can now be awarded based 
on this shortest path. Since the weight of each edge, which represents the cost of edit 
is non-negative, the shortest possible path may be found using Dijkstra's shortest path 
algorithm. Hence, given a set of edits and using a given set of canonicalizations, the 
shortest path in the graph, as defined above, gives the edit sequence with the least 
cost. We call the cost of the edits as the \textit{weighted edit sequence distance}.

\textbf{Theorem 1}: In the space of edits considered by our system and in the 
given space of canonicalization the edit sequence with least cost from the student query 
to a query that is canonically equivalent to a given correct query can be found using a shortest path algorithm. 

\subsection{Shortest Path Algorithm}
\label{sec:edit_path:algo}

\begin{algorithm}
	\caption{: partialMarks($CQ$, $SQ$)}
	\label{algo:shortest}	
	\renewcommand{\algorithmicrequire}{\textbf{Inputs:}}
	\renewcommand{\algorithmicensure}{\textbf{Output:}}	
	\begin{algorithmic}[1]
		\REQUIRE $CQ$: Canonicalized correct query\\
		\hspace{0.75cm}$SQ$: Student query with syntactic canonicalizations done\\
		\ENSURE Awarded marks, Edit Sequence 	
		\STATE $EQ$ $\leftarrow$	\{\}
		\STATE $totMarks$ $\leftarrow$ getTotalMarks($CQ$)
		\STATE $EQ$.add($SQ$, $totMarks$, \{\})
		\WHILE{$EQ$.isNotEmpty()}	
			\STATE ($SQ$, $marks$, $editSeq$) $\leftarrow$ $EQ$.popMaxMarksQuery() 		
			\IF{isCanonicallyEquivalent($CQ$, $SQ$)}
				\RETURN ($marks/totMarks$, $editSeq$)
			\ENDIF					 
			\STATE $SQEdits$ $\leftarrow$ editsOfStudentQuery($SQ$,$CQ$) 
			\FOR{Each $(SQ_i,cost_i,edit_i) \in SQEdits$}
				\STATE $newMarks$ $\leftarrow$ marks - $cost_i$
				\IF{$newMarks\leq 0$}
					\STATE continue
				\ENDIF
				\STATE $newEditSeq$ $\leftarrow$ $editSeq$.add($edit_i$)
				\IF{$EQ$.contains($SQ_i$)}
					\IF{$EQ$.get($SQ_i$).$marks<newMarks$)}
						\STATE $EQ$.update($SQ_i,newMarks,newEditSeq$)
					\ENDIF
				\ELSE
					\STATE $EQ$.add($SQ_i,newMarks,newEditSeq$)
				\ENDIF
			\ENDFOR
		\ENDWHILE
		\RETURN (0,\{\})
	\end{algorithmic}
\end{algorithm}

The algorithm for partial marks using the shortest path is shown in 
Algorithm~\ref{algo:shortest}. The algorithm does not actually create the entire graph, 
but effectively generates parts of it as needed. The algorithm takes as input CQ - a 
canonicalized correct 
query and SQ - a student query with only syntactic canonicalizations applied.
The total marks for the query is computed based on the number of components in the 
correct query by the function getTotalMarks.
EQ is a set that stores triplets of edited student query, remaining marks and the list of 
edits made to reach the edited query. 
It is initialized by adding the original student query with total marks of the query as remaining marks and 
an empty edit set (Line 2).

As shown in Section~\ref{sec:canonical:marks}, canonicalization may increase the edit 
distance between queries since it does not take into account potential edits into 
account. Edits on a query may make some canonicalization rules inapplicable that were 
applicable on the original query. In Algorithm~\ref{algo:shortest}, we do not edit the 
correct query. Hence the correct query may be canonicalized upfront. 
In order to apply the edits on a student query, the syntax used in the student query 
should be comparable to the correct query. For e.g if the student query uses WITH clauses 
or BETWEEN predicates while the correct query does not, it may be difficult to compare 
the queries to find which guided edits should be applied. Hence, we apply syntactic 
canonicalizations to the student query before applying edits.

We iterate the loop till EQ is empty or we have found a match.
Since we are trying to find the shortest path, the query with the 
highest marks in the EQ set, SQ (i.e. the query with the lowest distance from the 
original student query) is chosen for the current iteration and removed from EQ (Line 4).
The function isCanonicallyEquivalent checks if the edited query is canonically equivalent 
to the correct query. If the queries are equivalent,
the algorithm terminates and returns the fraction of marks awarded for the student query. 
The fraction can be later multiplied by the total marks set by the instructor for the 
query.

If the queries do not match, guided edits of the student query are generated
along with the 
costs of each edit using the function editsOfStudentQuery.
\fullversion{These edits along with the cost of the edits are generated by comparing the 
student and correct query trees as described in Section~\ref{sec:top_bottom}.}
\short{\ns{The edits can be generated by comparing the student and query trees as 
described in \cite{arxiv}.}}
From the current marks, the cost for an edit $cost_i$ can be deducted to get the marks 
$newMarks$ for the edited query $SQ_i$. 
Any edited query that has $newMarks$ as 0 or less can be discarded; the student would 
get 0 or less marks if the current sequence of edits is considered. We assume that 
students do not get negative marks for an answer. 

If the edited query is 
already present in the set EQ with a lower mark than $newMarks$, the query is replaced 
in EQ since we have now found a shorter path to reach the edited query (Lines 15-18). If 
the edited query is not present in EQ, the edited query along with the $newMarks$ and the 
new edit sequence is added to EQ. 
If EQ is empty, it implies that all edited queries that could have given greater than 0 
marks have already been considered, 
hence 0 marks is returned for the query.

Note that, syntactic canonicalization may increase the edit distance for student 
queries that use a WITH clause. Consider a student query that 
defines one non-recursive WITH clause and the WITH clause has an error that requires one 
edit to make it correct. The WITH clause 
is used twice in the student query. Applying one edit to 
the WITH clause would make the query correct. However, if we first inline the WITH clause 
and then find edits, two edits would be required.\footnote{We thank Rahul Sharma for 
pointing this out.} Hence when applying edits, we should deduct marks based on the edits 
that would have been required on the original student query. In the space of edits 
we consider, other syntactic 
canonicalizations do not increase the edit distance.

\subsection{Greedy Heuristic Solution}
\label{sec:heuristic}
Since the shortest path algorithm considers multiple options at each step, it can be very 
expensive for queries with a large number of components. We propose, as an alternative, a 
greedy approach that uses a cost-benefit model. 
\fullversion{When generating edits of a student query, we 
consider all guided edits. We also compute the canonicalized edit distance
for each edited query as described in Section~\ref{sec:edit_distance}.

}
For each edit that is made to the student query, there is some 
benefit due to the reduction in the canonicalized edit distance. Each edit has a 
certain cost associated with it as described above. We compute the cost-benefit as 
$benefit-cost$ and use it to 
pick the best edit for the next step. This helps us prune edits that may 
not be beneficial; e.g. removing an extra node from the student query that may have 
been removed anyway because of canonicalization later.

The heuristic algorithm proceeds in a similar manner as the exhaustive algorithm shown in 
Algorithm~\ref{algo:shortest}. The key difference is that in the exhaustive algorithm all 
$SQ_i$ where  $newMarks$ is greater than 0 are retained. In the greedy algorithm, before 
the while loop ends we compute the cost-benefit for each edited query in EQ. We keep 
only the edited query with the best cost-benefit in EQ and discard the rest. It 
should be noted that the value of benefit for the best query could be 0 or even negative. 
We do not discard such queries and continue the algorithm. As a result, EQ has only one
entry at a time.

\fullversion{\subsection{System Details and Discussion}
\label{sec:system}

We have implemented the techniques described in this paper as part of the XData grading 
system. Test data generated by the XData system 
is used to check student SQL queries for correctness and if the query is found to be 
incorrect, edit distance based partial marking is used to assign partial marks to the 
student query. When setting questions for an assignment, the instructor can change the 
weights of different components of the query as desired. This allows instructors to give 
more importance to some query constructs over others.

Instructors can also set up assignments in learning mode. In this mode, the students can 
get immediate feedback indicating whether the query is correct or not. Earlier, the XData 
system used to show failed datasets in case the student query was incorrect. However, 
showing failed datasets may not be effective. Consider the case where a correct query 
has a selection condition $r.A>10$ while the student query has the selection conditions 
$r.A<10$. If the selection conditions in the queries are in conjunction, the student 
query would fail almost all the test cases and showing test cases may not be very useful. 
Showing edits to the students would be a more useful feedback mechanism and would let 
them know the errors they made in their query.
 
Student queries that fail on datasets generated based on correct queries are 
incorrect i.e. there are no false negatives. 
However, in case the student query has additional conditions, evaluating student queries 
using datasets generated based on correct queries may not be able to catch errors. For 
example, if the student query has an additional selection condition \smalltt{r.A$>$5}, 
this selection condition would not be considered when generating datasets and the error 
may not get caught.    
For catching such errors, datasets need to be 
generated on the student queries as well. However, dataset generation in an expensive 
process. Query canonicalization, on the other hand, 
can find students queries that match a correct query without giving false positives. 
Hence queries marked correct by canonicalization are correct.
Datasets generated based student queries that do not match a correct query 
based on canonicalization but match datasets generated based on correct queries 
can be used to check correctness of student queries.
As shown in Section~\ref{sec:expt:canonical}, in practice, there could be few student 
queries like these. This would improve accuracy for correctness without being very 
expensive.

Canonicalized student queries may be clustered by placing canonically equivalent queries 
in one cluster. Correct student queries from each cluster can help the instructor reduce 
the effort of adding correct queries in case a canonically equivalent set of queries is 
correct but does not have any instructor provided correct queries. 
}
\section{Related Work}
\label{sec:relwork}

To the best of our knowledge, there is no other system for awarding partial marks to
student SQL queries. Related work include the following. \\

\noindent \textit{SQL Query Equivalence}:
For a restricted class of conjunctive queries, techniques based on 
tableau~\cite{tableaux} 
and its 
extensions such as \cite{tableaux1}, \cite{tableaux2} can be used to determine query 
equivalence. We, on the other hand, target arbitrary SQL queries.
As shown in \cite{jayram}, SQL query equivalence is undecidable in general.

The Cosette prover~\cite{cosette} can be used to test for equivalence of SQL queries. 
Given two queries, Cosette can infer if the queries are equivalent, non-equivalent or 
whether Cosette is not able to prove either of the two. Chu et al. in 
\cite{semantic} use U-semiring to model SQL queries and check for query equivalence.  Our 
canonicalization system, 
besides checking for equivalence also gives a weighted edit distance which is useful as a 
heuristic when deciding the order of edits. Besides, unlike \cite{semantic}, 
\cite{cosette} we also handle query 
features like \smalltt{ORDER BY}, outer joins, nulls, strings and more types of 
subqueries. 

The Gradience system \cite{gradience} like our XData system, allows instructors to 
test SQL queries by comparing query results on test datasets. However, unlike XData,
Gradience does not generate the test datasets, which must be provided by the instructor,
and provides no support for partial marking. RAT{\small{EST}} \cite{ratest} provides 
feedback to 
students by deriving small datasets (from existing larger datasets) that produce 
different results in a student query as compared to a correct query. Tuples that 
are not part of existing larger datasets may not be used in RAT{\small{EST}} as a 
dataset. In addition to showing incorrect test cases, the XData system can also 
show students the minimal edits required to make their query correct. \\

\noindent \textit{Tree Edit Distance}:
Tree edit distance between a given pair of trees is a well-studied problem, with 
efficient dynamic programming techniques, as described in \cite{survey:treeedit}.
In our context, there may be many correct query structures, and many irrelevant
syntactic/semantic variations; our techniques address these issues by using
canonicalization and edit sequences.\\

\noindent \textit{Automated Grading and Feedback for Programming Assignments}:
CPSGrader~\cite{cpsgrader} considers automated grading and feedback generation to student 
submissions in assignments on cyber-physical systems. It uses constraint synthesis along 
with a number of reference solutions to provide feedback to incorrect student submissions.
AutoGrader~\cite{autograder} describes techniques to provide automated feedback for 
basic python programs where the specification is known and the errors are predictable. 
They use program synthesis using SKETCH and define a high-level error modeling language 
for providing correction rules to identify errors in student programs.
CLARA~\cite{gulwani_cluster}  uses dynamic program analysis to cluster correct 
student 
programs and provides repair steps for incorrect programs to the closest cluster.
SARFGEN~\cite{sar} generates student feedback for introductory level programming 
exercises. Given an incorrect student program, it searches for the 
closest reference program, from a given set of reference programs, based on the program 
control flow, and ``aligns'' the two programs using canonicalization 
rules. 
It then generates correction suggestions by finding the minimum edits to the student 
program to match the chosen reference program. 
In contrast, our system, works on a very different language, uses a large set of 
canonicalizations, finds a minimum edit sequence to reach any correct query, and uses the 
edit sequence to award partial marks.

Partial marking in the context of programming languages is more complicated because of 
the program control flow, variable reassignment. We look at simple tree structures of 
SQL queries along with the constraints on the database. We can use complex 
canonicalization rules based on SQL that can be useful in eliminating syntactically 
irrelevant differences between SQL queries. However, for grading SQL queries we 
also need to consider the constraints in the database which is not required for grading 
programs.
To the best of our knowledge, query editing to award partial marks or to give feedback to 
student SQL queries has not been addressed previously. 

\section{Experimental Results}
\label{sec:expt}

We conducted experiments to test the fairness of our techniques and the 
runtime for evaluating partial marks for the queries. The student SQL queries used in these experiments were taken from student 
submissions in an undergraduate database course offered at IIT Bombay from 2015 to 2017. 
\ns{We do not generate any incorrect query for the experiments.}
The queries used several SQL features including subqueries, outer joins, set 
operators and aggregates with grouping. \ns{The list of correct queries used 
in the experiments are provided in 
\fullversion{Appendix~\ref{sec:queries}}\short{\cite{arxiv}}.}

\subsection{Effectiveness of Equivalence Checking Using Canonicalization}
\label{sec:expt:canonical}
A subproblem of our partial marking scheme is the use of canonicalization to check for 
equivalence.
In the first experiment, we test the effectiveness of canonicalization to directly check 
equivalence on 
student SQL queries. 
\fullversion{The canonicalization techniques are applied on all student and 
correct queries. If the canonicalization techniques worked correctly, student queries 
which are semantically equivalent to a correct query should be marked as correct.} If a 
query is marked as correct by canonicalization, we know that the query is correct; false 
positives are not possible. However, in case a query is marked as incorrect, it could be 
because of cases that we did not canonicalize and hence false negatives are possible. 

We also checked queries for errors using test data generated by running XData on the 
instructor query only. If XData marks a student query as incorrect, it has found a 
dataset on which the student and instructor query do not match; false negatives are not 
possible. On the other hand, if the query is marked as correct it may be because datasets 
to catch certain errors were not generated. Hence false positives are possible. However, 
this technique was able to catch more errors than TAs for an earlier course as shown in 
\cite{xdata:vldbj15}. 

\begin{table}\setlength{\tabcolsep}{3pt}
    \centering
    \begin{tabular}{|c| c| c|c| c| c|c| c| c|c|c| } 
\hline
\fullversion{Q. & SQ & CQ & CQ & $C_{data}$ &$C_{canon.}$&  XData & Canon.\\
No. &  &  & Size &  & &  FP & Acc.(\%)\\}
\shorttab{Q. No. & SQ & CQ & CQ Size & $C_{data}$ & $C_{canon}$ & XData FP & Canon. Acc. 
(\%) \\ }
\hline
CQ1	    &114	&1 	&3		&112	&112	&0	&100\\ %
CQ2		&94	    &1	&5		&88		&88	&0	&100\\	%
CQ3		&111	&1 	&5		&86		&86	&0	&100\\	%
CQ4		&104	&2 	&9-10	&79		&74	&5	&100 \\	%
CQ5		&117	&2 	&10-11	&104	&95	&1	&92.2\\	%
CQ6		&95		&3 	&10-14	&79		&67	&1	& 85.9\\	%
CQ7		&116	&3 	&10-14	&105	&90	&1 	&86.5\\	%
CQ8		&95		&3 	&14-17	&51		&44	&0	&86.3\\	%
CQ9		&89		&2 	&18		&70		&61	&2 	& 89.7\\	%
CQ10	&115	&5 	&17-25	&102	&85	&0 	& 83.3\\	%
CQ11 	&87 	&2 	&25-27 	&58		&55 	&0	& 94.8 \\	%
CQ12	&93		&3 	&29		&34		&29	&0 	& 85.3\\	%
CQ13 	& 92 	&1 	& 35 	& 58 	&51	&0 	& 87.9\\	%
CQ14	&88		&2 	&18-32	&52		&48 	& 1	&94.1\\	%
CQ15	&79		&4 	&41-50	&45	&28		&12 	& 84.8\\%
\hline
\hline
Total &1489 &-& - & 1123 & 1013 & 23 & 92.1\\
\hline
\end{tabular} 
\caption{Effectiveness of Canonicalization}
 \label{tab:canonicalization}
\end{table}

The result of the experiment is shown in Table~\ref{tab:canonicalization}. 
The column SQ shows the number of student queries being evaluated, CQ shows 
the number of correct queries used for the evaluation. One measure of the complexity of a 
query is the number of nodes in the flattened tree as shown 
in column CQSize. The column $C_{data}$ shows the number of queries marked 
correct by datasets generated by XData, while $C_{canon.}$ shows the number of queries 
marked correct because of canonicalization. 

For queries that were marked as correct by 
XData but were found non-matching by canonicalization, we manually evaluated the queries 
to check if they were correct or not. The number of student queries that were marked 
correct by dataset but were not actually correct i.e the false positives for XData is 
shown in the column XDataFP. 

The accuracy of canonicalization can be obtained 
based on  the fraction of correct student queries that were marked equivalent to the 
correct query by canonicalization. This accuracy (Canon.Acc. as shown in the table) is 
computed as $C_{canon.}/(C_{data}-XDataFP)$

The overall accuracy of canonicalization in our experiments is 92.1\%. We found that, in 
particular, our canonicalization missed cases where student queries had additional 
query components that did not have any effect on the query results. For example, if the 
correct query was \smalltt{CQ} some students wrote queries like (i) \smalltt{CQ UNION Q'} 
where \smalltt{Q'} is a query that returns empty results or is equivalent to 
\smalltt{CQ} (ii) \smalltt{CQ INTERSECT Q'}, 
where \smalltt{Q'} is equivalent to the correct query. Removing such extraneous query 
components, that do not affect the query result, during canonicalization is part of 
future work.

The overall accuracy is satisfactory for our purpose.
For some small fraction of queries, partial marks may be lower due to the use of 
canonicalization for equivalence checking, since more edits may be needed before the 
edited query is found to be canonically equivalent to a correct query. 

\subsection{Checking Fairness}
In this experiment, our goal is to find the fairness of the partial marks given by our 
weighted edit sequence distance using the greedy heuristic. We cannot directly compare 
the partial marks given our algorithm with partial marks given for student assignments in 
earlier years. Partial marks for previous years 
were given using different techniques - canonicalization and manual grading by TA. Also, 
we were not aware of the grading scheme used by the TAs including which errors in the 
query were penalized more relative to others. Most importantly, assigning partial marks 
manually is very difficult, and 
grades given are only approximate and not necessarily consistent. Hence making a direct 
comparison between 
manual partial marking and partial marks generated by our system is not desirable. 

\begin{table}\setlength{\tabcolsep}{1pt}
    \centering
       
    \begin{tabular}{|c| c| c|c|c|c| c| c|c|}
\hline
Q.	&SQ 	&CQ	&CQ		&Matches	&Accuracy		&Matches	&Accuracy \\
No.	&Pairs	& 	&Size	&Canon.	&Canon.(\%)	&Edit		&Edit(\%)\\
\hline
CQ3		& 6		&1 	&5		&6	&100	&6	&100\\
CQ4		& 12	&2 	&9-10	&10	&83.3	&12&100\\
CQ5		& 11	&2 	&10-11	&11	&100	&11&100\\
CQ6		& 21	&3 	&10-14	&16	&76.2	&19&90.5\\
CQ7		& 17	&3 	&10-14	&17	&100	&17&100\\
CQ8		& 20	&3 	&14-17	&16	&80		&18&90\\
CQ9		& 16	&2 	&18		&10	&62.5	&15&93.8\\
CQ10	& 25	&5 	&17-25	&17	&68		&20&80\\
CQ11	& 21	&2 	&25-27 	&10	&47.6	&18& 85.7\\
CQ12	& 20	&3 	&29		&15	&75		&17&85\\
CQ13	& 23	&1 	& 35 	&7	&30.4	& 23&100\\
CQ14	& 6		&2 	&18-32	&6	&100	&6&100\\
CQ15	& 30	&4 	&41-50		&9	&30		&29&96.7\\
\hline
\hline
Total & 228 &-	&-		&150&65.8&211&92.5\\
\hline
\end{tabular}
 \caption{Evaluation of Grading Fairness}
        \label{tab:partial}
\end{table}

Instead, we judged the fairness of our techniques as follows.
We awarded partial marks using weighted edit sequence distance to each incorrect 
query. For each assignment question, we created random pairs of incorrect student 
queries. We provided these query pairs to two volunteers\footnote{
\ns{One of the volunteers was a TA for the undergraduate database course at IIT Bombay in 
2018 while another had been an instructor for databases courses at another institute.}}
 (without giving them the partial 
marks awarded using our techniques) 
and asked the volunteers to classify the query pairs into three buckets (a) The first 
query should get more marks (b) The second query should get more marks (c) Both queries 
should get almost the same marks even though they may have different errors. 
We then classified the query pairs in the above 3 categories using the partial marks 
awarded by query edits. If the partial marks differed by less than 10\% we classified the 
query as being almost equal. \ns{Only incorrect student queries can be used in this 
experiment since partial marking is not applicable to correct queries.} \fullversion{For 
CQ1 and CQ2, the errors in the student query were similar and hence we could not generate 
meaningful pairs of incorrect student queries.}

In Section~\ref{sec:canonical:marks}, we discussed why partial marks awarded based on 
canonicalized edit distance would not be fair. In this experiment, we 
also evaluated the effectiveness of partial marks by using the canonicalized edit 
distance. As discussed in 
Section~\ref{sec:edit_distance}, 
we call each part of a query such as selection, projection, aggregate or subquery a 
component. In order to compute the marks using canonicalized edit distance, we used the 
formula 
\begin{equation*}
\frac{max(0,  (\Sigma_{c \in \textrm{qc}} W_c*N_c) -(\Sigma_{c \in 
\textrm{qc}} W_c*E_c ))}{\Sigma_{c \in \textrm{qc}} W_c*N_c}*maxMarks
\end{equation*}
where $qc$ is the set of all query components, $N_c$ is the number of nodes of the 
component in the correct query under consideration, $W_c$ is the weight assigned to a 
component and $E_c$ is the edit distance for the component.
Similar to edit sequence based partial marks, we classify the same 
student query pairs into buckets using marks awarded based on the 
given formula.

The result of the experiment is shown in Table~\ref{tab:partial}. The column SQPairs 
shows the number of incorrect student query pairs that we considered. CQ shows the number 
of correct queries used to evaluate student assignments. CQSize shows the number of 
nodes present in the instructor query and gives some measure of the complexity of 
the correct queries. 
The column Matches Canon indicates the number of student query pairs that 
were added to the same bucket by canonicalization as well as by 
the volunteers. Accuracy Canon. gives the corresponding accuracy which is computed as 
Matched Canon./SQPairs.
Similarly, the column Matches Edit indicates the number of student query pairs that 
were added to the same bucket by our edit sequence based partial marking as well as by 
the volunteers, and Accuracy Edit gives the accuracy.

While partial marking based on canonicalized edit distance works well for simpler 
queries, 
it performs poorly for more complex queries; the overall accuracy is 65.8\%. 
Our edit based partial marking system works much better and its overall accuracy is 
92.5\%. 
In several cases, we found that a few edits enabled other canonicalizations that made the 
edited query equivalent to the correct query. Such edits appear to have matched human 
intuition. 
For some cases, our canonicalization techniques converted OUTER JOINs to INNER JOINs and 
removed redundant relations from student queries thus not penalizing their use. The 
volunteers considered the use of outer joins/additional relations as significant errors 
even though they were technically equivalent. Hence there was a difference in the 
buckets in which the 
volunteers placed the pairs with the bucket classification  as per our techniques; 
turning off outer joins to inner join canonicalization and redundant relation removal 
would be an option to model human intuition on the degree of error. \\ 

\noindent\textbf{Real World Usage:}
We used the grading system to grade student queries in database courses conducted at IIT 
Bombay in Autumn 2018. 
We graded over 1800 student queries 
automatically including awarding partial marks to incorrect student queries using our 
edit based partial marking technique. Except for a few queries that used 
constructs like \smalltt{RANK}, \smalltt{PARTITION}, string functions and 
expressions, we were able to handle all other queries. Implementation for these is an 
area of future work. Once the assignments were graded, students were allowed to contest 
the marks that they had obtained. Students only contested for 4 query submissions of 
which 2 were genuine. 
In both cases, we found implementation bugs in our code because of which students were 
not awarded partial marks fairly.  In contrast, in earlier years, anecdotally, when 
partial marks were 
awarded manually, many students had contested their marks.

%\subsection{Comparison with Edit Distance Marks}

\subsection{Comparison of Exhaustive and Heuristic Algorithms}
In order to measure the effectiveness of our heuristic techniques compared to the 
exhaustive approach, we compared the running time and marks awarded to incorrect student 
queries using both approaches. Queries marked as incorrect using datasets generated by 
XData are used to compute partial marks.  
This experiment was run on a computer with an Intel(R) Core(TM) i7-3770 3.40GHz CPU, and 
16 GB of memory, running Ubuntu Linux. 

\begin{table} \setlength{\tabcolsep}{4pt}
\centering
   
    \begin{tabular}{|c|c|c|c|c|c|c| }
\hline
\fullversion{Q. & SQ & CQ & CQ & Match & $T_{greedy}$ & $T_{exhaustive}$\\
No.&  &  & Size & (\%) & (in ms) & (in ms)\\}
\shorttab{Q. No. & SQ & CQ & CQ Size&Match (\%) & $T_{greedy}$ (in ms) & $T_{exhaustive}$ 
(in ms)\\}
\hline
 CQ1	& 2 	& 1 	& 3 		&100	& 171 	& 3776\\
CQ2 	& 6 	& 1 	& 5  		&100	& 165 	& 171\\
CQ3 	& 25 	& 1 	& 5 		&100	& 200	& 210\\
CQ4 	& 25 	& 2 	& 9-10 		&100	& 200 	& 549\\
CQ5 	& 13 	& 2 	& 10-11 	&100	& 219	& 293\\
CQ6 	& 16	& 3 	& 10-14 	&100	& 246 	& 2355\\
CQ7 	& 11 	& 3 	& 10-14 	&100	& 252 	& 1869 \\
CQ8 	& 44 	& 3 	& 14-17 	&100	& 199 	& 1172\\
CQ9 	& 19 	& 2 	& 18 		&100	&  253 	& 445\\
CQ10 	& 13 	& 5 	& 17-25 	&100*	&  305 	& 7187*\\
CQ11	& 29 	& 2 	& 25-27 		&100	&  378 	& 535\\
CQ12 	& 59 	& 3 	& 29 		&100*	&  278 	& 31014*\\
CQ13 	& 34 	& 1 	& 35    	&100*	&  334 	& 31872*\\
CQ14 	& 36 	& 2 	& 18-32 	&100*	&  245 	& 7006*\\
CQ15 	& 34 	& 4 	& 41-50 		&100*	&  246 	& 12672*\\
\hline
%\hline
%Total & 70 & - & - & 253 & -\\
%\hline
\end{tabular}
 \caption{Heuristic vs. Exhaustive}
    \label{tab:greedy}
\end{table}

The results are shown in Table~\ref{tab:greedy}. SQ and CQ are 
respectively the number of student queries and correct queries considered during 
evaluation. 
Similar to the previous two tables, CQSize gives the number of components in the correct 
queries. $Match$ shows the percent of cases for which the marks 
awarded by the exhaustive and heuristic greedy algorithm matched. 
For several cases, the exhaustive 
technique ran out of memory, even with the memory limit set to 14 GB in Java. For such 
cases, we computed results excluding such student queries and the numbers are marked with 
an asterisk. 

The partial marks awarded in both cases 
were identical for all student queries that could be evaluated by both techniques. 
$T_{greedy}$ gives the average time taken to evaluate a student query 
against one correct query using the greedy algorithm while $T_{exhaustive}$ gives the 
average time taken to evaluate a student query against one correct query using the 
exhaustive algorithm.

As shown in the table the exhaustive algorithm takes much longer. 
In particular, for cases where the student query had a large number of errors (and hence 
got awarded low  marks), the exhaustive approach had 
to explore a much larger set of paths compared to the greedy approach.

From the above experiment, we conclude that the heuristic algorithm is accurate when 
evaluating real student submissions and performs much better, in terms of time taken, as 
compared to the exhaustive algorithm while not running out of memory for any query.

\section{Conclusion}
In this paper, we discussed techniques for evaluating student queries and awarding 
partial marks to incorrect student queries. Our system is useful for automated evaluation 
of student SQL queries and would benefit database instructors and TAs. The experimental 
results show that our techniques work well in practice.
Our partial marking scheme was used successfully in 2018 for courses at IIT Bombay and 
IIT Dharwad. 
The source code and binaries available for download from 
\url{http://www.cse.iitb.ac.in/infolab/xdata}. 

Areas of future work include adding more canonicalization rules like unnesting of
subqueries and support for more SQL features such as windowing, ranking and OLAP 
features. 

\noindent\textbf{Acknowledgments:}
We thank Bharath Radhakrishnan for the initial implementation of the 
canonicalization code. We 
also thank Ravishankar Karanam and Aarti Sharma for evaluating student 
queries in the second experiment.

\fullversion{
\section*{APPENDIX}

\appendix
\section{Syntactic Canonicalization}
\label{sec:canonical_basic}

\ns{For the purpose of brevity, we illustrate our canonicalization rules via examples, 
but our implementation has carefully coded rules checked manually for correctness.}
 We consider the following syntactic rewrite rules  
include the following.
\begin{itemize}[leftmargin=*]
 \item \textbf{Attribute disambiguation}: An attribute \smalltt{A} without a relation 
is changed to \smalltt{r.A} where \smalltt{A} is inferred to be from \smalltt{R}. \ns{We 
assign an id to each instance of a relation. When comparing attributes between queries, 
we use the relation instance id and the attribute name from the underlying relation. 
This allows us to handle cases where queries rename attributes or where the FROM clause 
subquery has an alias.}

\item \textbf{WITH Clause Elimination}:
Non-recursive WITH clauses in the query are replaced in the query by expanding the WITH 
clauses inline.

\item\textbf{BETWEEN Predicate Elimination}: 
BETWEEN predicates are replaced with the equivalent conditions 
using the relational operators. For example, \smalltt{r.A BETWEEN 5 and 10} is replaced 
with \smalltt{r.A>=5  AND r.A<=10}.

\item\textbf{Normalization of Relational Predicates}: 
Selection conditions involving NOT are converted to remove the NOT operator by 
adjusting the relational operator appropriately. For example, \smalltt{NOT(A>B)} is 
converted to \smalltt{A<=B}. 
Relational predicates involving $>$ (resp. $>=$) are converted to $<$ (resp. $<=$), by 
exchanging the operands; for example, \smalltt{A>B} is converted to \smalltt{B<A}.
In case of an INNER JOIN, the selection conditions are combined with the join 
conditions as shown in Figure~\ref{fig:flattenedTree}.

\item\textbf{Normalization of Nested Queries}: 
A nested subquery with an IN/SOME connective is converted to use
an EXISTS connective, by using the attributes involved in the
IN/SOME connective to create a correlation condition.
For example, 
\\ \smalltt{r.A >SOME (SELECT s.A FROM s WHERE s.B$>$10)}
\\ is converted 
to \\ \smalltt{EXISTS (SELECT s.A FROM s WHERE s.B$>$10 AND r.A>s.A)}\\
Similarly nested subqueries with ALL/NOT IN is converted to NOT EXISTS with an 
appropriate correlation condition. \ns{ When converting to NOT EXISTS, if an attribute 
used in the connective is nullable, we add appropriate \smalltt{IS NULL} conditions 
as discussed in \cite{subquery}.}
\eat{For example, \\ \smalltt{r.A >ALL (SELECT s.A FROM 
s)}\\ is converted to \\ \smalltt{NOT EXISTS (SELECT s.A FROM s WHERE 
\\\hspace*{0.5cm}r.A<=s.A 
OR r.A IS NULL OR s.A IS NULL)}}

DISTINCT clauses are deleted from EXISTS/ IN/ ALL/ SOME subqueries, as well as
their NOT variants, regardless of the presence of duplicates.

\item \textbf{Join Processing}: 
Any NATURAL INNER JOIN is replaced with an INNER JOIN with equivalent join conditions 
added using the ON clause. 
Occurrences of USING clause in join conditions is replaced with 
ON clause with the equivalent join conditions. Any expression using RIGHT OUTER JOIN is 
converted to equivalent expression using LEFT OUTER JOIN. 

\item \textbf{ORDER BY in subqueries}: ORDER BY clauses without LIMIT that are part of a 
subquery do not affect the query result and are hence removed. 

\item \textbf{Flattening operators}:
For any operator $op$ that is commutative and associative, expressions of the form 
$(A~op~B)~op~C$ can be written as $op(A, B, C)$ if it is possible to add the 
operators at individual $op$ nodes to the $op$ node of the transformed tree. Multiple 
instances of the operator can be replaced by a single flattened operator.
Inner joins without DISTINCT, aggregation and GROUP BY are flattened to ensure join 
orders do not affect the comparison. Similarly UNION (ALL), INTERSECT (ALL) as well as  
predicates involving AND or OR are flattened. 

\end{itemize}

\section{Semantic Canonicalization}
\label{sec:semantic}

We consider the following semantic canonicalizations.
\subsection{Canonicalizing DISTINCT}

Duplicate removal using SELECT DISTINCT can be redundant if there are no duplicates
in the list of attributes; if we infer the absence of duplicates, the DISTINCT clause are 
removed.
Primary key constraints on input relations, coupled with equality predicates in select 
and join predicates can be used to infer the absence of duplicates in the result of joins,
as described in \cite{paulley:cascon93}. Similarly, for INTERSECT ALL absence of 
duplicates in at least of the inputs, and for
EXCEPT ALL in the left input, means we drop the ALL clause.

 Removing duplicates from FROM clause subqueries depends on the JOIN.
In case there are DISTINCT clauses in the FROM clause subquery and the query does not 
contain aggregates, we consider the following cases
\begin{itemize}
\item In case there is a DISTINCT above the FROM clause subquery in the query tree, the 
DISTINCT is removed from the FROM clause.
\item In case the primary keys of all relations are present in the final result, the 
DISTINCT clause in all FROM clauses are removed.
\end{itemize}

Consider the query
\smalltt{\\\tab SELECT A, B, S.pk\\
\tab FROM S, (SELECT DISTINCT A,B FROM R) WHERE pred\\
} where \smalltt{S.pk} is the primary key of \smalltt{S}.

Since the primary key of S is output along with distinct values of the FROM clause 
subquery, the join above the FROM clause subquery will not generate any duplicates in the 
query results.
The DISTINCT can be moved from the subquery to the outer query; the query could be 
written as  
\smalltt{\\\tab SELECT DISTINCT A, B, S.pk\\
\tab FROM S, (SELECT A,B FROM R) WHERE pred
} 

In case there is a DISTINCT in a FROM clause subquery and we infer that the join above 
the FROM clause does not create duplicates (if each of the remaining relations in the 
FROM clause have unique attributes in the projection of the outer query SELECT clause), 
we pull up the 
DISTINCT above the subquery to the outer query. 
The removal of DISTINCT in the FROM clause subquery may enable flattening, as is this 
case in the above example.

\subsection{Join Canonicalization}

Removal of redundant joins, and conversion of outer joins to inner joins 
are well-known steps in query optimization.  We use them as 
part of our canonicalization.
Consider the following query:
\\\smalltt{
\tab SELECT student.id, department.dept\_name\\ 
\tab FROM student  INNER JOIN department USING(dept\_name)
}\\
If \smalltt{student.dept\_name} is non nullable, and is a
foreign key referring to \smalltt{department.dept\_name}. 
The non-nullable foreign key dependency ensures that for
each employee tuple $t_1$ there exists a matching
department tuple $t_2$ (i.e., one s.t. $t_1[$dept\_name$]$ $=$ $t_2[$dept\_name$]$).
Since the projection attribute, \smalltt{department.dept\_name}
can be replaced by \smalltt{student.dept\_name} (since the two attributes have the same 
value due to the join condition), the 
query is rewritten as the equivalent query: 
\\\smalltt{
\tab SELECT student.id, student.dept\_name FROM student
}

Consider the query:
\\\smalltt{
\tab SELECT * \\\tab FROM department LEFT OUTER JOIN student USING(dept\_name)\\
\tab WHERE student.dept\_name = 'Biology'
}

The selection condition, \smalltt{student.dept\_name = 'Biology'}, fails when 
\smalltt{student.dept\_name} has a null value.  
Thus the query is equivalent to the one where an INNER JOIN is used.
In general, if at a point in the query above a LEFT OUTER JOIN,
there is a null-rejecting condition on an attribute from the right 
input of the LEFT OUTER JOIN, we replace the LEFT OUTER JOIN 
by an INNER JOIN. 

In case of a LEFT OUTER JOIN where (i) the joining attributes include all foreign key 
references from the left operand to the right operand and (ii) the foreign keys are non 
nullable, the LEFT OUTER JOIN can be converted to an INNER JOIN. Similar is the case with 
a RIGHT OUTER JOIN. 
For the query, 
\\\smalltt{
\tab SELECT * \\\tab FROM student LEFT OUTER JOIN department USING(dept\_name)\\
}
if \smalltt{student.dept\_name} is non nullable, and is a
foreign key referring to \smalltt{department.dept\_name}, the LEFT OUTER JOIN is 
converted 
to an INNER JOIN.

\subsection{Predicate Canonicalization}

Query predicates are pushed to the lowest possible level in the flattened query tree.  
In case of inner joins predicates of both inputs of the join are pushed down the query 
tree. 
In case of left outer join only the predicates based on left input are pushed down, 
similarly for right outer joins only predicates based on right input are pushed down. 
This is a common technique used in query optimizers.

Selection conditions involving \smalltt{A<B} are converted to \smalltt{A<=B+1}, 
provided both operands are of integer type. 

Consider the following query
\\\smalltt{
\tab SELECT student.dept\_name \\
\tab FROM  student INNER JOIN department USING(dept\_name)
\\\tab WHERE student.dept\_name LIKE 'English\%'
}

In this query, \smalltt{SELECT department.dept\_name}
can be used in place of \smalltt{SELECT student.dept\_name}, 
since the two attributes are guaranteed to have the same value
thanks to the join condition, \smalltt{student.dept\_name $=$ department.dept\_name}. 

In general, when $A=B$, $B=C$, $C=D$ .., are conjunct predicates of a query, 
attributes $A$, $B$, $C$, $D$, ... are said to belong to the same \textit{equivalence 
	class}; any occurrence of an attribute in an equivalence class can be replaced with 
any other attribute from the equivalence class, at any place in the query tree that is 
above the 
occurrence of the join conditions, without changing the result of the query.

A canonicalization step is therefore performed by replacing all occurrences 
of an attribute above join condition, by the lexicographically
least attribute from its equivalence class.
In the above query, since \smalltt{department.dept\_name} lexicographically precedes 
\smalltt{student.dept\_name}, \smalltt{student.dept\_name} is replaced by 
\\\smalltt{department.dept\_name} in the SELECT clause.
In case there are equality conditions involving constants, the constant is treated as the 
lexicographically least attribute.

Mapping variables to equivalence classes is used by query optimizers for join reordering 
and correct selection estimation whereas we use it for comparing queries.

\subsection{Order By and Group By Canonicalization} 

Functional dependencies can be used to infer that textually different 
ORDER BY or GROUP BY clauses are actually equivalent \cite{neumann:vldb04},
which is used for query optimization. 

In addition to the functional dependencies of the query, we also consider additional 
functional dependencies that can be inferred like those based on equivalence classes. For 
example, if there are two equivalence class $(A,B)$ and $(C,D)$, and there is a 
functional dependency $A\rightarrow C$, we also consider the functional 
dependencies $A\rightarrow D$, $B\rightarrow C $ and $B\rightarrow D$.

Consider an SQL query $Q$ with the clause ORDER BY $a, b$. 
Let us suppose that $Q$ satisfies the functional dependency $a \rightarrow b$, then 
$Q$  is equivalent to a query $Q'$ obtained by replacing 
the ordering clause with ORDER BY $a$. 
Due to the functional dependency, two tuples with the same value for $a$ would have the
same value for $b$, making the ordering by $b$ irrelevant.
ORDER BY clauses are canonicalized by removing all attributes that are
functionally determined by other attributes appearing earlier in the ORDER BY clause.

Consider the following query 
\\\smalltt{
\tab SELECT id, COUNT(*) \\ 
\tab FROM student INNER JOIN takes USING(id) \\
\tab GROUP BY id, name
}

Suppose \smalltt{id} functionally determines \smalltt{name} (for example, because
\smalltt{id} is the primary key of the student relation).
Then, the GROUP BY clause can be equivalently written as \smalltt{GROUP BY id}. 

However, unlike with ORDER BY clauses, there may be completely different sets of 
attributes that give the same grouping, and getting a unique canonicalization is not 
possible as shown in \cite{neumann:vldb04}. This may happen when a relation or a join 
result has more 
than one super key. 
Consider two group by clauses to be \smalltt{GROUP BY a,b} and \smalltt{GROUP BY a,c}. If 
\smalltt{\{a,b\} $\rightarrow$ c} and \smalltt{\{a,c\} $\rightarrow$ b}, then the GROUP 
BY clauses are equivalent 
even though the attributes do not match.
Hence as stated in \cite{neumann:vldb04}, we add all attributes that can be determined by 
the group by attributes using functional dependencies.

\section{Queries}
\label{sec:queries}

\begin{enumerate}[label=CQ \arabic*)]

\item \smalltt{SELECT id, name FROM student}

\item \smalltt{SELECT * FROM student WHERE name < 'Mahesh'}

\item \smalltt{SELECT building FROM classroom
WHERE capacity > 50}

\item \begin{enumerate}[label=\alph*)]
	\item \smalltt{SELECT name FROM student, department
	\\WHERE student.dept\_name = department.dept\_name \\\tab AND building = 
	'KReSIT'}
	\item \smalltt{SELECT name FROM student
	\\WHERE dept\_name IN (SELECT dept\_name FROM department \\\tab WHERE building 
	= 'KReSIT')}
\end{enumerate}

\item \begin{enumerate}[label=\alph*)]
	\item \smalltt{SELECT DISTINCT s.id, s.name \\FROM student s, takes t
	\\WHERE s.id = t.id AND t.grade = 'F'}
	\item \smalltt{SELECT id, name FROM student
	\\WHERE id IN (SELECT id FROM takes \\\tab WHERE grade = 'F')}
\end{enumerate}

\item \begin{enumerate}[label=\alph*)]
	\item \smalltt{SELECT id, name FROM student s
	\\ WHERE NOT EXISTS (SELECT id FROM takes \\\tab WHERE grade = 'F' AND s.id = 
	takes.id)}
	
	\item \smalltt{WITH result as \\\tab (SELECT s.id, s.name, semester, grade 
	\\\tab FROM student s LEFT OUTER JOIN takes t 
	\\\tab  ON (s.id = t.id AND t.grade = 'F'))
    \\SELECT id, name FROM result \\ WHERE semester IS NULL}
	
	\item \smalltt{SELECT s.id, s.name FROM student s
	\\EXCEPT\\
	SELECT s.id, s.name FROM student s, takes t
	\\ WHERE s.id = t.id AND t.grade = 'F'}
\end{enumerate}

\item \begin{enumerate}[label=\alph*)]
	\item \smalltt{SELECT id, name FROM student
	\\ WHERE id NOT IN (SELECT id FROM takes \\\tab WHERE grade = 'F')}
	
	\item \smalltt{WITH result as \\\tab (SELECT s.id, s.name, semester, grade 
	\\\tab FROM student s LEFT OUTER JOIN takes t 
	\\\tab  ON (s.id = t.id AND t.grade = 'F'))
    \\SELECT id, name FROM result \\ WHERE semester IS NULL}
	
	\item \smalltt{SELECT s.id, s.name FROM student s
	\\EXCEPT\\
	SELECT s.id, s.name FROM student s, takes t
	\\ WHERE s.id = t.id AND t.grade = 'F'}
\end{enumerate}

\item \begin{enumerate}[label=\alph*)]
	\item \smalltt{SELECT id, name FROM student s 
	\\WHERE tot\_cred > 50 AND
	NOT EXISTS (SELECT * \\\tab FROM takes  \\\tab WHERE takes.id = s.id AND takes.grade 
	= 'A')}
	
	\item \smalltt{(SELECT id, name FROM student \\\tab WHERE tot\_cred > 50)
	\\EXCEPT\\
	(SELECT id, name FROM student NATURAL JOIN takes WHERE grade = 'A') }
	
	\item \smalltt{SELECT id,name FROM student 
	\\WHERE id IN ((SELECT id 
		\\\tab FROM student WHERE tot\_cred > 50 ) 
		\\\tab EXCEPT  
		\\\tab (SELECT id \\\tab\tab FROM takes WHERE grade = 'A'))}
\end{enumerate}

\item \begin{enumerate}[label=\alph*)]
	\item \smalltt{SELECT course\_id, title FROM course
	\\EXCEPT\\
	SELECT course.course\_id, course.title \\\tab FROM course, section, 
	time\_slot
	\\\tab WHERE course.course\_id = section.course\_id
	\\\tab AND section.time\_slot\_id = time\_slot.time\_slot\_id \\\tab AND start\_hr < 
	12}
	
	\item \smalltt{SELECT course.course\_id, course.title FROM course
	\\WHERE course.course\_id NOT IN \\\tab (SELECT course.course\_id 
	\\\tab FROM course, section, time\_slot
	\\\tab WHERE section.time\_slot\_id = time\_slot.time\_slot\_id
	\\\tab AND course.course\_id = section.course\_id \\\tab AND time\_slot.start\_hr < 
	12)}
\end{enumerate}

\item \begin{enumerate}[label=\alph*)]
	\item \smalltt{SELECT s.course\_id,c.title,s.year,s.semester,s.sec\_id 
	\\FROM section s, course c, department d
	\\WHERE s.course\_id = c.course\_id \\AND s.building <> d.building
	\\AND c.dept\_name = d.dept\_name}
	
	\item \smalltt{SELECT course\_id,title,year,semester,sec\_id 
	\\FROM (course NATURAL JOIN section) as cs \\ WHERE building <>
	(SELECT building FROM department \\\tab WHERE department.dept\_name = cs.dept\_name)}
	
	\item \smalltt{WITH \\\tab course\_name AS (SELECT course\_id, title FROM course), 
	\\\tab	info(course\_id, title, year, semester, sec\_id, building) \\\tab\tab AS
	(SELECT s.course\_id, c.title, s.year, s.semester, \\\tab\tab\tab s.sec\_id, 
	s.building 
	\\\tab\tab FROM section s INNER JOIN course\_name AS c 
	\\\tab\tab ON c.course\_id = s.course\_id)
	\\ SELECT i.course\_id,i.title i.year,i.semester,i.sec\_id \\FROM info 
	i, department d, course c 
	\\WHERE c.course\_id = i.course\_id \\AND d.dept\_name = c.dept\_name \\ AND 
	d.building <> i.building}
	
	\item \smalltt{(SELECT course\_id,title,year,semester,sec\_id \\\tab FROM section 
	NATURAL 
	JOIN course) \\EXCEPT \\
	(SELECT course\_id,title,year,semester,sec\_id \\ FROM
	(SELECT * FROM course NATURAL JOIN section) P \\\tab NATURAL JOIN department)}
	
	\item \smalltt{SELECT c.course\_id,c.title,s.year,s.semester,s.sec\_id 
	\\ FROM course c, section s	
	\\WHERE c.course\_id = s.course\_id AND s.building
	\\\tab NOT IN ( SELECT d.building \\\tab FROM department d,course ci
	\\\tab WHERE ci.course\_id = c.course\_id \\\tab AND ci.dept\_name = 
	d.dept\_name)}
\end{enumerate}

\item \begin{enumerate}[label=\alph*)]
	\item \smalltt{SELECT id, name FROM instructor
	\\EXCEPT \\
	SELECT i.id, i.name \\\tab FROM instructor i, teaches te, takes ta
	\\\tab WHERE i.id = te.id AND te.course\_id = ta.course\_id
	\\\tab AND te.year = ta.year AND te.semester = ta.semester \\\tab AND 
	te.sec\_id = ta.sec\_id AND ta.grade = 'A'}
	
	\item \smalltt{SELECT id, name FROM instructor i
	\\ WHERE NOT EXISTS (SELECT t2.id \\\tab FROM takes ta, teaches te
	\\\tab WHERE ta.course\_id = te.course\_id \\\tab AND ta.sec\_id = te.sec\_id AND 
	ta.semester = te.semester \\\tab AND ta.year = te.year AND ta.grade = 'A' 
	\\\tab AND i.id = te.id)}
\end{enumerate}

\item \begin{enumerate}[label=\alph*)]
	\item \smalltt{SELECT DISTINCT ts.day 
	\\FROM teaches t, section s, time\_slot ts
	\\ WHERE t.course\_id = s.course\_id \\AND t.semester = 
	s.semester
	\\AND t.year = s.year AND t.sec\_id = s.sec\_id
	\\AND s.time\_slot\_id = ts.time\_slot\_id \\AND s.semester = 'Fall' 
	AND s.year = '2009' \\AND t.id = '22222'}
	
	\item \smalltt{SELECT day FROM time\_slot 
	\\WHERE EXISTS (SELECT * \\\tab FROM teaches t JOIN 
	section s \\\tab\tab USING
	(course\_id,sec\_id, semester, year)
	\\\tab WHERE t.id='2222' AND s.time\_slot\_id=ts.time\_slot\_id \\\tab AND
	t.semester = 'Fall' AND t.year = '2009')}
	
	\item \smalltt{SELECT day FROM time\_slot  \\WHERE time\_slot\_id
	IN (SELECT time\_slot\_id \\\tab FROM section WHERE course\_id IN (SELECT course\_id 
	\\\tab\tab FROM teaches
	WHERE id = '22222' \\\tab\tab AND semester = 'Fall' AND year = '2009'))}
\end{enumerate}

\item \begin{enumerate}[label=\alph*)]
	\item \smalltt{(SELECT s.id, name \\\tab FROM student s, takes t, course c 
	\\\tab WHERE s.id = t.id AND
	t.course\_id = c.course\_id \\\tab AND c.dept\_name = 'Comp. Sci.' AND year < 2010)
	\\INTERSECT\\
	(SELECT s.id, name \\\tab FROM student s, takes t, course c
	\\\tab WHERE s.id = t.id AND
	t.course\_id = c.course\_id 
	\\\tab AND c.dept\_name = 'Comp. Sci.' AND year> 2010)}	
\end{enumerate}

\item \begin{enumerate}[label=\alph*)]
	\item \smalltt{SELECT DISTINCT c.course\_id, c.title \\FROM course c, section s, 
	time\_slot t
	\\WHERE c.course\_id = s.course\_id
	\\AND s.time\_slot\_id = t.time\_slot\_id
	\\AND t.end hr >= 12 AND c.dept\_name = 'Comp. Sci.'}
	
	\item \smalltt{(SELECT c.course\_id, c.title 
	\\\tab FROM course c, section s, time\_slot t
	\\\tab WHERE c.dept\_name = 'Comp. Sci.' \\\tab AND c.course\_id = s.course\_id
	\\\tab AND s.time\_slot\_id = t.time\_slot\_id)
	\\EXCEPT\\
	(SELECT c.course\_id,c.title \\\tab FROM course c, section s, time\_slot t
	\\\tab WHERE c.dept\_name = 'Comp. Sci.' \\\tab AND c.course\_id = s.course\_id
	\\\tab AND s.time\_slot\_id = t.time\_slot\_id \\\tab AND t.end\_hr < '12')}
\end{enumerate}

\item \begin{enumerate}[label=\alph*)]
	\item \smalltt{(SELECT i.id, i.name 
	\\\tab FROM instructor i, teaches te, takes ta
	\\\tab WHERE i.id = te.id AND te.course\_id = ta.course\_id
	\\\tab AND te.year = ta.year AND te.semester = ta.semester \\\tab AND 
	te.sec\_id = ta.sec\_id AND ta.grade IS NOT NULL)
	\\EXCEPT\\
	(SELECT i.id, i.name 
	\\\tab FROM instructor i, teaches te, takes ta
	\\\tab WHERE i.id = te.id AND te.course\_id = ta.course\_id
	\\\tab AND te.year = ta.year AND te.semester = ta.semester \\\tab AND 
	te.sec\_id = ta.sec\_id AND ta.grade = 'A')}
	
	\item \smalltt{SELECT i.id, i.name 
	\\FROM instructor i, teaches te, takes ta
	\\WHERE i.id = te.id AND te.course\_id = ta.course\_id
	\\AND te.year = ta.year AND te.semester = ta.semester 
	\\AND te.sec\_id = ta.sec\_id AND ta.grade IS NOT NULL 
	\\AND i.id NOT IN (SELECT i1.id \\\tab FROM instructor i1, teaches te1, takes ta1
	\\\tab WHERE i1.id = te1.id AND te1.course\_id = ta1.course\_id
	\\\tab AND te1.year = ta1.year AND te1.semester = ta1.semester
	\\\tab AND te1.sec\_id = ta1.sec\_id AND ta1.grade = 'A')}

	\item \smalltt{SELECT i.id, i.name 
	FROM instructor i
	\\WHERE instructor.id NOT IN (SELECT te.id 
	\\\tab FROM teaches te, takes ta
	\\\tab WHERE te.course\_id = ta.course\_id \\\tab AND te.sec\_id = ta.sec\_id
	AND te.semester = ta.semester \\\tab AND te.year = ta.year AND ta.grade = 'A')
	\\AND i.id IN (SELECT te.id 
	FROM teaches te, takes ta \\\tab WHERE te.course\_id = ta.course\_id
	\\\tab AND te.sec\_id = ta.sec\_id AND te.semester = te.semester
	\\\tab AND te.year = ta.year AND ta.grade IS NOT NULL)}

	\item \smalltt{(SELECT instructor.id, instructor.name FROM instructor
	\\\tab EXCEPT \\\tab 
	SELECT i.id, i.name 
	\\\tab\tab FROM instructor i NATURAL JOIN teaches te
	\\\tab\tab JOIN takes ta ON (ta.course\_id = te.course\_id
	\\\tab\tab AND ta.sec\_id = te.sec\_id AND ta.semester = te.semester 
	\\\tab\tab AND ta.year = te.year)
	\\\tab\tab WHERE takes.grade = 'A')
	\\INTERSECT\\
	(SELECT i.id, i.name 
	\\\tab\tab FROM instructor i NATURAL JOIN teaches te 
	\\\tab\tab JOIN takes ta ON (ta.course\_id = te.course\_id
	\\\tab\tab AND ta.sec\_id = te.sec\_id AND ta.semester = te.semester
	\\\tab\tab AND ta.year = te.year) \\\tab\tab WHERE ta.grade IS NOT NULL)}
\end{enumerate}

\end{enumerate}

\bibliographystyle{ACM-Reference-Format}
}

\short{
\bibliographystyle{splncs04}
}

\bibliography{references}
\balance

\end{document}